\documentclass[%
 reprint,
 amsmath,amssymb,
 aps,
]{revtex4-2}

\usepackage{graphicx}
\usepackage{dcolumn}
\usepackage{bm}
\usepackage{textcomp}
\usepackage{multirow}
\usepackage[colorlinks,
            linkcolor=blue,
            anchorcolor=blue,
            citecolor=blue]{hyperref}

\begin{document}

\preprint{APS/123-QED}

\title{Response of Comagnetometer to Exotic Spin-Dependent Couplings}

\author{Yixin Zhao}
\author{Xiang Peng}
\author{Teng Wu}%
 \email{wuteng@pku.edu.cn}
 \author{Hong Guo}
\email{hongguo@pku.edu.cn}
\affiliation{%
State Key Laboratory of Photonics and Communications, School of Electronics,\\
and Center for Quantum Information Technology, Peking University, Beijing 100871, China
}

\date{\today}

\begin{abstract}
Comagnetometers offer unique advantages in measuring exotic spin-dependent interactions by suppressing magnetic noise.
Understanding the frequency response to exotic interactions is essential for evaluating and improving the sensitivity of comagnetometers, as the signals often exhibit distinct spectral features.
We conduct theoretical and experimental studies on the frequency response of a differential comagnetometer. 
We derive analytical expressions for the frequency response of the differential comagnetometer to both magnetic field and exotic couplings, along with the conversion relationship between them.
Based on this, we revise the method for determining the axion–nucleon coupling strength with comagnetometers.
Furthermore, we introduce a light-shift-based calibration protocol to verify the response to exotic interactions.
This work helps predict the sensitivity of comagnetometers in detecting exotic fields and advances the exploration of dark matter.
\end{abstract}

\maketitle

\section{\label{sec:level1}Introduction}
A magnetometer is a sensor used to measure magnetic fields, characterized by a finite bandwidth and a frequency-dependent response~\cite{ripka2002magnetic,bertoldi2005magnetoresistive}. 
Its frequency response determines how effectively it can reconstruct external magnetic or magnetism-like signals across a range of frequencies. 
In recent years, several studies have explored the dynamic response of atomic magnetometers to time-dependent fields~\cite{bevilacqua2016multichannel,bevilacqua2021spin,zhang2020frequency,li2020continuous,zhang2025influence}.
Exotic non-magnetic couplings can induce shifts in atomic energy levels and alter spin precession frequencies~\cite{abel2017search,safronova2018search, jackson2023probing}, enabling magnetometers to detect not only magnetic fields but also spin-dependent anomalous interactions.
The capability offers a promising approach for investigating physics beyond the Standard Model and contributes to multi-messenger strategies in the quest to understand the nature of dark matter~\cite{abbott2017multi,dailey2021quantum}.
Building on these advances, researchers have increasingly employed magnetometers to explore new physics in table-top experiments, broadening the scope from detecting magnetic interactions to exotic spin-dependent couplings~\cite{afach2021search, jiang2021search,nichol2022elementary}.

A magnetometer serves as a powerful tool for exploring new physics, but the performance can be compromised by fluctuations in ambient magnetic fields. 
A comagnetometer typically compares signals from multiple atomic species in the same volume, effectively canceling common-mode magnetic noise and maintaining sensitivity to exotic interactions~\cite{venema1992search, terrano2021comagnetometer}.
In the past, comagnetometers have been primarily employed for detecting static signals, such as spin-gravity coupling~\cite{venema1992search,jackson2017constraints, ShengDong2023SG} and electric dipole moments (EDMs)~\cite{rosenberry2001atomic,baker2006improved}, with little consideration given to the response to oscillating signals or other exotic signals with frequency components.
Dark matter and spin-dependent interactions may exhibit a wide range of frequency characteristics, manifesting either as broadband signals (e.g., domain walls~\cite{afach2021search,roberts2017search}) or as coherent oscillations at specific frequencies (e.g., axion stars~\cite{jackson2018searching} and axion wind~\cite{graham2013new, gao2022axion}).
Therefore, a thorough understanding of the frequency response of comagnetometers is crucial for accurately evaluating and enhancing its detection capabilities.

Comagnetometers can be generally categorized into differential types~\cite{tullney2013constraints,jackson2017constraints,wang2020single,zhang2023search} and self-compensating types~\cite{kornack2002dynamics,lee2018improved}, based on the operating principles.
The former separates magnetic and non-magnetic contributions to atomic energy level shifts by comparing the outputs of different magnetometers subjected to the same magnetic field~\cite{venema1992search,tullney2013constraints,jackson2017constraints}.
The latter relies on the coupling between alkali-metal atoms and nuclear spins. 
By tuning the bias magnetic field to the self-compensation point, it suppresses low-frequency magnetic noise while preserving sensitivity to exotic interactions~\cite{kornack2002dynamics,lee2018improved}.
Recent studies have explored the frequency response of self-compensating comagnetometers to such interactions~\cite{padniuk2022response,padniuk2024universal,rosenzweig2024atomic}.
In contrast, the response characteristics of a differential comagnetometer to exotic interactions have not yet been investigated.

This work presents a theoretical and experimental investigation of the frequency response of a differential comagnetometer.
Using a perturbative approach, we derive analytical expressions for its response to both oscillating magnetic and non-magnetic signals, along with their conversion relation, enabling predictions of non-magnetic responses based on its magnetic response. 
We also revise the calculation method for determining the axion–nucleon
coupling strength with comagnetometers.
Experimentally, we measure the frequency response of a free induction decay (FID) comagnetometer and perform simulations based on theoretical calculations to interpret the experimental results. 
In addition, we introduce an experimental protocol that uses light shifts as calibration pulses to test the consistency between theoretical predictions and measured responses to non-magnetic interactions.
This work enables the interpretation of comagnetometer signals in terms of exotic interaction strengths and supports accurate measurements of frequency-dependent exotic interactions.

\section{theoretical analysis}\label{se2}
In this paper, we define the amplitude frequency response of the sensor as the ratio of the output to input signal amplitudes at a given frequency.
The comagnetometer we focus on in this work is primarily a single-species atomic comagnetometer, though the principles discussed can also be applied to other differential comagnetometers.
A single-species atomic comagnetometer is a differential comagnetometer that utilizes two hyperfine levels of an alkali-metal atom~\cite{wang2020single,zhao2024suppression}. 
By eliminating systematic errors arising from magnetic-field gradients caused by spatial separation between different atomic species, it offers a promising platform for fundamental physics.
First, based on previous calculations of magnetic field response~\cite{bevilacqua2021spin,zhang2020frequency,li2020continuous}, we calculate the response of the two hyperfine levels to the magnetic field separately, with the coupling between the magnetic field and the atomic nucleus taken into account.
At the same time, we consider the presence of anomalous fields and calculate the frequency response of the magnetometer to exotic field-electron, exotic field-proton, and exotic field-neutron couplings.
Based on the above calculations, we discuss the frequency response of a differential comagnetometer to both magnetic and non-magnetic fields.

\subsection{Frequency response of a magnetometer to magnetic and exotic spin-dependent perturbations}

To compare the response of a conventional magnetometer to magnetic and non-magnetic perturbations, we express equivalent magnetic fields corresponding to interactions of the exotic field with electrons, protons, and neutrons as $\mathbf{b}_{\mathbf{e}}$, $\mathbf{b}_{\mathbf{p}}$,
$\mathbf{b}_{\mathbf{n}}$.
Taking the Bell-Bloom magnetometer as an example, considering the electron polarization relaxation rate $R_{e}$ and optical pumping rate $R_{p}(t)$, the evolution equation for the electron polarization $\mathbf{P}$ of alkali-metal atoms with nuclear spin $I$ can be written as 
\begin{equation}
\begin{aligned}
    \frac{\mathrm{d} \mathbf{P}}{\mathrm{~d} t}=& \pm \frac{\gamma_e}{2 I+1}\left(\mathbf{B}+\mathbf{b}_{\mathbf{e}}\right) \times \mathbf{P}\\
    +&\gamma_e\left(1 \mp \frac{1}{2 I+1}\right)\left(\frac{g_I}{g_S}\mathbf{B}+\mathbf{b}_{\mathbf{p}}+\mathbf{b}_{\mathbf{n}}\right) \times \mathbf{P}\\
    -&R_e \mathbf{P}-R_p(t) (\mathbf{P}-\mathbf{P}_{0}). 
\end{aligned}\label{eq:dP1}
\end{equation}
Here, $\gamma_{e}=\mu_{B}g_{S}/\hbar$ is the gyromagnetic ratio of electron. $\mathbf{P}_{0}$ is the equilibrium spin polarization if there is no relaxation and magnetic field.
$R_{p}(t)\approx R_{p}(\cos (\omega t)+1)/2$, $\omega$ is the modulation frequency of the amplitude of the pump light.
``$\pm$" signs correspond to the two hyperfine levels $F_{a}=I+J$ and $F_{b}=I-J$ (for ground-state alkali-metal atoms, $J = 1/2$), respectively. 
The detailed derivation of Eq.~\eqref{eq:dP1} can be found in Appendix~\ref{app:dP}.

Assume a bias magnetic field $B_{0}$ is applied along the $z$-axis, and the pump light propagates along the $x$-axis, i.e., $\mathbf{P}_{0}=P_{0}\hat{x}$.
The perturbation field is aligned with the $z$-axis (here we focus only on the anomalous variations along the quantum axis), with the expression given by
\begin{equation}
    \mathbf{B}_{s}(t)=B_{s}\cos(\omega_s t)\hat{z}, \label{eq:perturbation}
\end{equation}
where $B_{s}$ is the magnitude of the perturbation field. 
Let $\gamma=\gamma_{e}/(2I+1)$ denotes the gyromagnetic ratio of the alkali-metal atoms, neglecting the hyperfine structure. 
$\gamma B_{s}/R_{e}$ is a small quantity.

\subsubsection{Magnetic perturbation}
\begin{figure*}[t]
\includegraphics[width=1\linewidth]{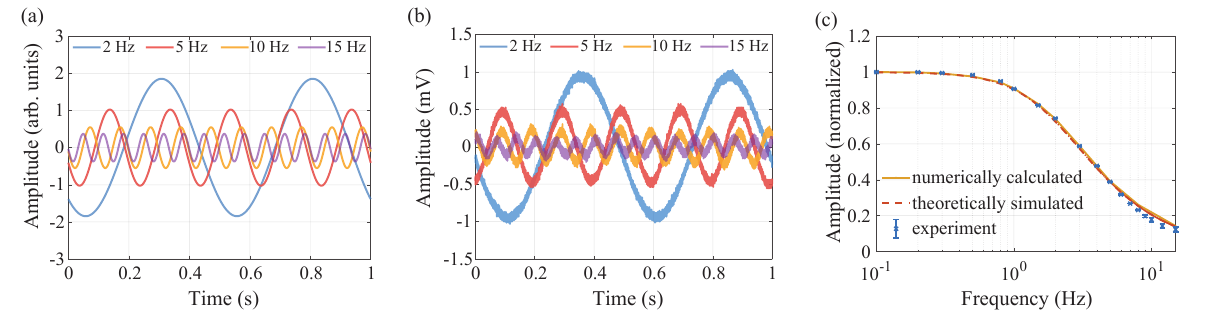}
\caption{\label{figure1} The frequency response of a magnetometer to magnetic field perturbations. (a) Simulated demodulated magnetometer signal under magnetic field perturbations, obtained via numerical integration of the Bloch equations.
(b) Experimentally measured demodulated magnetometer signal under magnetic field perturbations.
In (a) and (b), colored curves represent responses to perturbations at different frequencies. The difference in the phase response is due to the fact that the initial time of signal acquisition in the experiment is random.
(c) The normalized amplitude response of the magnetometer to perturbations at different frequencies. 
The yellow curve represents the simulation results obtained by numerically integrating the Bloch equations, and the red dashed line represents the approximate analytical solution derived using perturbation expansion.
The blue ``$\times$" represents experimental results, the error bar represents the standard error calculated from three measurements.}
\end{figure*}

Substituting Eq.~\eqref{eq:perturbation} into Eq.~\eqref{eq:dP1}, $\mathbf{B}\rightarrow \mathbf{B}+\mathbf{B_{s}}(t)$, $\mathbf{b_{e}}=\mathbf{b_{p}}=\mathbf{b_{n}}=0$, let $\gamma_{a}=\gamma(1+2I\cdot \frac{g_{I}}{g_{S}})$, the evolution of $\mathbf{P}$ for $F_a$ can be rewritten as
\begin{equation}
 \frac{\mathrm{d} \mathbf{P}}{\mathrm{~d} t}=\gamma_a\left(\mathbf{B}+\mathbf{B}_{s}(t)\right) \times \mathbf{P}-R_e \mathbf{P}-R_p(t) (\mathbf{P}-\mathbf{P}_{0}). \label{eq:dPv} 
\end{equation}
We can use numerical integration methods, specifically ordinary differential equations (ODEs), to simulate the response of atoms to magnetic perturbations. 
The magnetometer’s response to oscillating fields at different frequencies obtained through the method is shown in Fig.~\ref{figure1}(a).
Additionally, we can provide analytical solution using the perturbative expansion approach, with the detailed derivation given in Appendix~\ref{app:perturbation}.

For $F_{a}=I+1/2$, when demodulating the electron polarization signal at $\omega=\gamma_a B_0$, the quadrature component is ideally zero without a perturbation field. 
Calculations show that when a perturbation field at frequency $\omega_s$ is present, the quadrature component $S_{\text{Qu},a}^{(1)}$ oscillates at frequency $\omega_{s}$, with an amplitude depending on $\omega_{s}$,
\begin{equation}
    S_{\text{Qu},a}^{(1)}  =\frac{\gamma_{a} B_s R_{p,a}P_0}{R_{e,a} \sqrt{\omega_s^2+R_{e,a}^2}} \sin(\omega_{s}t+\varphi),\label{eq:LIA}
\end{equation}
$\varphi$ represents the initial phase. It can be seen that the output of a magnetometer to magnetic field perturbations is similar to that of a low-pass filter. 
For $F_{b}=I-1/2$ , the output signal can be obtained through similar calculations.
Let $\gamma_{b}=-\left[1-(2I+2)\cdot\frac{g_{I}}{g_{S}}\right]\gamma$, when $\omega=\gamma_{b}B_{0}$,
\begin{equation}
    S_{\text{Qu},b}^{(1)}  =\frac{\gamma_{b} B_s R_{p,b}P_0}{R_{e,b} \sqrt{\omega_s^2+R_{e,b}^2}} \sin(\omega_{s}t+\varphi) .
\end{equation}
$R_{p,j}$ and $R_{e,j}$ ($j=a,b$) denote the light pumping rate and the relaxation rate for different hyperfine levels.
In the experiment, we measure the variation of the demodulated signal from a single magnetometer under magnetic perturbations, as shown in Fig.~\ref{figure1}(b). The detailed experimental procedure is provided in Appendix~\ref{app:exp}.

Once the variation of the demodulated signal with respect to the perturbation field frequency is known, the voltage signal can be converted into a magnetic field signal by the slope of the magnetic resonance curve. 
This enables the derivation of the magnetometer’s magnetic field response, which relates the measured field amplitude to the frequency variation of the perturbation field.
According to Eq.~\eqref{eq:P0}, the slope of the magnetic resonance signal can be expressed as
\begin{equation}
    \frac{\text{d}V}{\text{d}\omega}=\frac{R_{p}P_{0}}{R_{e}^2}.\label{eq:dV}
\end{equation}
Therefore, the voltage signal can be converted into a frequency signal based on the slope and subsequently into a magnetic field. 
The relationship between the measured magnetic field variation amplitude $\delta B_{j}$ and the actual applied magnetic field magnitude $B_{s}$, which we define as the normalized field response, is given by
\begin{equation}
    \frac{\delta B_{j}}{B_{s}}=\frac{1}{|\gamma_{j}|}\frac{R_{e,j}^2}{R_{p,j}P_{0}B_{s}}|S_{\text{Qu},j}^{(1)}|=\frac{R_{e,j}}{\sqrt{\omega_{s}^2+R_{e,j}^2}},\label{eq:NFR}
\end{equation}
where $|S_{\text{Qu},j}^{(1)}|$ denotes the amplitude of $S_{\text{Qu},j}^{(1)}$.
For convenience, we directly use $\gamma_j$ to denote the absolute values hereafter.
The normalized field responses obtained from numerical calculations, theoretical simulations, and experimental measurements are shown in Fig~\ref{figure1}(c).
The $R_{e}$ used in the simulation is $2\pi\times 2.1$ Hz.
When $\omega_{s}=0$, i.e.,  the magnetometer detects a static magnetic signal, the measured signal matches the applied field exactly, with $\delta B_{j}/B_{s}=1$. 
As the signal frequency $\omega_s$ increases, the normalized field response gradually decreases and eventually approaches zero, indicating that the magnetometer becomes insensitive to high-frequency components. 

\subsubsection{Exotic spin-dependent perturbation}
Following a similar procedure as for magnetic perturbations mentioned above, by substituting $\mathbf{b_e}$, $\mathbf{b_p}$ or $\mathbf{b_n}$ in Eq.~\eqref{eq:dP1} with $\mathbf{B_s}$, the demodulated signal for different couplings can be expressed accordingly.
Utilizing the slope of the magnetic resonance signal in Eq.~\eqref{eq:dV}, the demodulated signal induced by anomalous interactions can be converted into an equivalent magnetic field.
The ratio between the field reconstructed by the magnetometer and the applied equivalent perturbation field is given by 
\begin{equation}
    k_{i}(\omega_{s})=\frac{\delta B_{i}}{B_s}=K_{i}\frac{R_{e,i}}{\sqrt{\omega_{s}^2+R_{e,i}^2}},\label{eq:ki}
\end{equation}
where $K_{i}$ for different couplings is summarized in Table~\ref{tab:table1}. 
From Eq.~\eqref{eq:ki} and Table~\ref{tab:table1}, we can see that for perturbations of the same magnitude caused by a magnetic field and an equivalent anomalous field, the magnetometer's response function follows the same line shape, with possible variations in the coefficients.
With the conclusion, the magnetometer signal enables estimation of the exotic interaction strength.
If the signal amplitude measured at a certain frequency $\omega_{s}$ is $A$, the equivalent magnetic field magnitude induced by a possible anomalous interaction can be inferred using the response coefficient as $A/k_i(\omega_s)$.
Once the equivalent magnetic field magnitude is obtained, the coupling constants for different anomalous interactions can be derived using Eqs.~([\eqref{eq:H1},~\eqref{eq:chi_I},~\eqref{eq:psedufield}]).

\begin{table}[b]
\renewcommand{\arraystretch}{1.3}
\caption{\label{tab:table1}%
  Response coefficients of a magnetometer to different interactions. 
}
\begin{ruledtabular}
\begin{tabular}{ccc}
\textrm{Interactions}&
\textrm{Angular momentum $F$ }&\textrm{Coefficients $K_{i}$}\\
\colrule
\multirow{2}*{$\mathbf{B}-\mathbf{S}\&\mathbf{I}$} & $F_a=I+1/2$ & 1  \\
~&$F_b=I-1/2$ &1\\
\colrule
\multirow{2}*{$\mathbf{\Xi}-\mathbf{S}$} & $F_a=I+1/2$ & $\gamma/\gamma_a$ \\
~&$F_b=I-1/2$ &$\gamma/\gamma_b$\\
\colrule
\multirow{2}*{$\mathbf{\Xi}-\mathbf{I}$} & $F_a=I+1/2$ & $2I\cdot\gamma/\gamma_a$ \\
~&$F_b=I-1/2$ &$(2I+2)\cdot\gamma/\gamma_b$\\
\end{tabular}
\end{ruledtabular}
\end{table}

\subsection{Frequency response of a comagnetometer to magnetic and exotic spin-dependent perturbations}~\label{subs:2B}

\begin{figure}[t]
\includegraphics[width=0.9\linewidth]{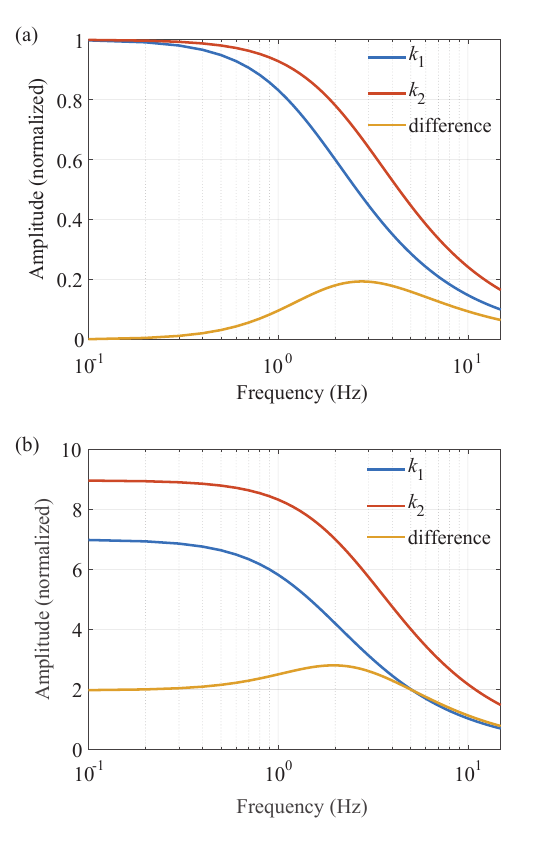}
\caption{\label{figure2} Simulated frequency responses of the two channels of the comagnetometer and the differential signal to magnetic and non-magnetic perturbations.(a) Responses of the two channels of the comagnetometer and the differential signal to magnetic perturbations. (b) Responses of the two channels of the comagnetometer and the differential signal to exotic field and nuclear spin coupling.}
\end{figure}

For a magnetometer, whether operating in open-loop or closed-loop mode, it is necessary to first convert the measured physical signal into a magnetic field using some conversion factors or transfer functions. 
However, a comagnetometer, similar to a gradiometer, is designed to measure the difference in field values between two magnetometers and to analyze its correlation with the original field~\cite{sheng2017microfabricated, sulai2019characterizing, dong2023recent}.
We first convert the measured signals from the two magnetometers into magnetic field, 
and then analyze the response of the differential signal to the applied field.
Using the results obtained from the differential signal, the output of the comagnetometer is defined as
\begin{equation}
\begin{aligned}
    S_{\text{Co}}(\omega_{s})&=\delta B_{1}(\omega_{s})-\delta B_{2}(\omega_{s})\\
    &=\left(k_{1}(\omega_{s})-k_{2}(\omega_{s})\right)\times B_{s}.\label{eq:Cooutput}
\end{aligned}
\end{equation}
If a comagnetometer is constructed using the two hyperfine levels of alkali-metal atoms, the output signal of the comagnetometer can be expressed by substituting Eq.~\eqref{eq:ki} and Table~\ref{tab:table1} into Eq.~\eqref{eq:Cooutput}.
The ratio of output to input signal amplitude is defined as the amplitude response of the comagnetometer,
\begin{equation}
    k_{\text{Co}}(\omega_{s})=\frac{|S_{\text{Co}}|}{B_{s}}=\left|K_{a}\frac{R_{e,a}}{\sqrt{\omega_{s}^2+R_{e,a}^2}}-K_{b}\frac{R_{e,b}}{\sqrt{\omega_{s}^2+R_{e,b}^2}}\right|.\label{eq:kco}
\end{equation}
For real magnetic field perturbations, $K_a=K_b=1$, the output is zero at low frequencies. 
As the frequency increases, if the relaxation rates of the two hyperfine levels are perfectly identical, the output remains zero, indicating that the comagnetometer is insensitive to external magnetic field perturbations. 
In general, due to differences in spin-exchange collisions cross-sections and power broadening, the relaxation rates of the two hyperfine levels exhibit slight discrepancies.
The discrepancy causes a non-zero differential signal at higher frequencies $\omega_s$, which in turn degrades the common-mode rejection. 
When the signal frequency approaches the bandwidth of the magnetometer, its frequency response drops below 1, gradually losing the ability to measure the magnetic field, and the differential signal of the comagnetometer approaches zero, as shown in Fig.~\ref{figure2}(a).

From Eq.~\eqref{eq:kco} and Table~\ref{tab:table1}, we can see that the comagnetometer's response to exotic field-electron interactions is similar to that of real magnetic fields.
Therefore, the main emphasis is placed on analyzing the comagnetometer’s response to exotic field-nucleon interactions.
Taking the exotic field-proton interactions as an example, for the same exotic field $\Xi$, the two hyperfine levels of the cesium atom experience the same effective pseudo-magnetic field $\mathbf{b_p}$, but the two magnetometers respond to the same $\mathbf{b_p}$ with different response coefficients.
Substituting the response coefficients for cesium, $K_{a}=2I\cdot\gamma/\gamma_a\approx7$ and $K_{b}=(2I+2)\cdot\gamma/\gamma_{b}\approx9$, into Eq.~\eqref{eq:kco}, we can derive the ratio between the output signal of the comagnetometer and the original signal.
At low frequencies, the ratio is approximately 2, indicating that the comagnetometer maintains its response to the exotic field-proton interaction.
As the frequency increases, the ratio shows a slight upward trend initially and eventually approaches zero, as shown in Fig.~\ref{figure2}(b).

Comparing Fig.~\ref{figure2}(a) and Fig.~\ref{figure2}(b), it can be observed that the frequency responses of the magnetometer to magnetic and non-magnetic fields differ only in amplitude and the ratio is not frequency-dependent in terms of signal shape, i.e., $k_{\Xi}/k_{B}$ is a fixed value.
As a result, once the magnetometer's frequency response amplitude to the magnetic field is known, the measured results can be converted to the response amplitude for the non-magnetic field using the conversion factor.
However, for the comagnetometer, due to the involvement of the differential signal between two channels, the conversion coefficient $k_{\text{Co},\Xi}/k_{\text{Co},B}$ for the frequency response to different interactions becomes frequency-dependent.
Therefore, it is essential to consider the frequency response characteristics of the comagnetometer at each frequency point when measuring magnetic field perturbations to accurately infer the strengths of the exotic couplings from the signal amplitude measured by the comagnetometer.

Based on the above theory, the measurement results of the comagnetometer can be converted into exotic spin-dependent coupling strengths through the following steps. 
(1) Transform the differential signal $S_{\text{Co}}$ of the comagnetometer into the frequency domain, and assume that its amplitude at a certain frequency $\omega_s$ is $A$, which may correspond to exotic field–proton coupling.
(2) Measure the responses of the two channels of the comagnetometer to real magnetic field to extract the relaxation rates 
$R_{e,a}$ and $R_{e,b}$ of the two hyperfine levels.
(3) The experimentally obtained values of $R_{e,a}$ and $R_{e,b}$, combined with $K_{a}$ and $K_{b}$ from Table~\ref{tab:table1}, are substituted into Eq.~\eqref{eq:kco} to obtain the ratio $k_{\text{Co}}$.
Next, $k_{\text{Co}}$ and the measured amplitude  $A$ (i.e., $ S_{\text{Co}})$ are used to calculate the equivalent magnetic field magnitude $B_{s}$ corresponding to the exotic coupling.
(4) Use Eq.~\eqref{eq:psedufield}, the equivalent magnetic field magnitude is converted into the exotic coupling strength. 
In this way, the comagnetometer measurements can be more accurately translated into the coupling strength with the frequency response properly accounted for.
Similarly, if no typical signal is detected at $\omega_s$, the coupling strength can be bounded based on the corresponding noise intensity.

\subsection{Correction to the axion–proton coupling strength based on comagnetometer response}

Take the axion wind as an example, if a signal is detected at a certain frequency $\omega_s$ using the comagnetometer, with a measured amplitude $A$, then based on the known frequency response of the comagnetometer, the properties of the axion field can be determined. 
Axions (or axion-like particles) are promising candidates for dark matter~\cite{peccei1977constraints}.
As the solar system moves through the cosmic axion background, it experiences a classical oscillating field commonly referred to as the ``axion wind"~\cite{graham2013new, gao2022axion}.
In natural units ($\hbar=c=1$), the axion field can be expressed as $a(t)=a_0\cos(m_a t+\phi)$, where $a_0$ represents the amplitude of the oscillating field, $m_a$ is the mass of the axion, and $\phi$ denotes the initial phase of the axion field.
According to the Schmidt model~\cite{schmidt1937magnetischen}, for Cs, the nuclear spin originates from the valence nucleon spin and its orbital angular momentum, the contribution of the proton spin to the overall nuclear spin is calculated to be $\sigma_p=-1/9$~\cite{kimball2015nuclear}.
The axion field-proton coupling Hamiltonian can be expressed as~\cite{graham2013new,stadnik2014axion}
\begin{equation}
\begin{aligned}
    H_{\text{wind}}&=\sigma_p\chi_p\mathbf{\nabla}a(\mathbf{r},t)\cdot\mathbf{I}\\
    &=-\frac{1}{9}\chi_{p}m_{a}a_{0}\sin(m_{a}t+\phi)\mathbf{v}\cdot\mathbf{I},
\end{aligned}
\end{equation}
where $\mathbf{v}$ is the relative velocity between the axion field and the Earth. Assuming that the energy density of the axion field accounts for the local dark matter density, $\rho_{\text{DM}}\approx0.4 \ \text{GeV/cm}^{3}$, and using the relation $\rho_{\text{DM}}\approx(1/2)m_{a}^{2}a_{0}^{2}$, one can estimate $m_{a}a_{0}$.
Considering the projection of the axion field-proton coupling along the direction of the bias magnetic field, according to Appendix~\ref{app:dP}, the resulting effective pseudomagnetic field is
\begin{equation}
    \mathbf{b_{p}}=-\frac{1}{9}\frac{\chi_{p}\sqrt{2\rho_{\text{DM}}}\sin(m_{a}t+\phi)v_{z}}{\mu_{B}g_{S}}\hat{z}.
\end{equation}
$b_{p}$ is the projection of the amplitude of $\mathbf{b_{p}}$ along the $z$-direction.
For axion field-proton couplings, the ratio of the signal amplitude measured by the comagnetometer to that of the original effective pseudomagnetic field is 
\begin{equation}
    k_{\text{Co,wind}}=\frac{A}{b_{p}}\approx \left|\frac{7R_{e,a}}{\sqrt{\omega_{s}^2+R_{e,a}^2}}-\frac{9R_{e,b}}{\sqrt{\omega_{s}^2+R_{e,b}^2}}\right|.\label{eq:kwind}
\end{equation}
Based on the procedure in Section~\ref{subs:2B} for extracting the coupling strength from the comagnetometer measurements, we can determine the axion–proton coupling strength through the following steps.
(1) Get the experimentally measured signal amplitude $A$ at frequency $\omega_{s}$.(2) Measure the response of the comagnetometer to magnetic field and obtain $R_{e,a}$ and  $R_{e,b}$. 
(3) Using Eq.~\eqref{eq:kwind}, calculate $k_{\text{Co, wind}}$ with $R_{e,a}$, $R_{e,b}$ and $\omega_s$. And then use $k_{\text{Co, wind}}$ and $A$ to determine the magnitude of the effective pseudomagnetic field $b_{p}$.
(4) Derive the axion field-proton coupling strength $\chi_p$.
As $\omega_s\rightarrow0$, $k_{\text{Co}}\approx2$, the relation between the signal amplitude and $\chi_p$ is given by
\begin{equation}
    A\approx\frac{\chi_{p}\sqrt{2\rho_{\text{DM}}}v_{z}}{36\gamma_{a}}.\label{eq:chip}
\end{equation}
The coupling strength formulas from earlier comagnetometers~\cite{wu2019search,bevington2020dual}, which neglect frequency response, take a form similar to Eq.~\eqref{eq:chip}.
However, as the oscillation frequency increases, applying Eq.~\eqref{eq:chip} without modification leads to considerable errors.
For example, with the simulation parameters used in Fig.~\ref{figure2}~($R_{e,a}=2\pi\times1.5$ Hz, $R_{e,b}=2\pi\times2.5$ Hz), the correction factor $k_{\text{Co}}\approx2.8$ at 2 Hz, meaning that the neglecting the frequency response results in an overestimation of $\chi_{p}$ by 40\%. 
At 20 Hz, $k_\text{Co}$ drops to 0.5, causing $\chi_{p}$ to be underestimated by 75\% if the frequency response is not taken into account.
Hertz-level oscillations correspond to axion-like particle masses around $10^{-15}$ eV, which is in a range that draws significant attention in the search for axion-like dark matter.

Therefore, if an exotic signal is present, neglecting the frequency response can lead to inaccurate estimation of the coupling strength.
If no exotic signal is observed, using experimental results to constrain the parameter space without accounting for the frequency response may exclude false parameter regions at higher frequencies, meaning that the absence of a dark matter signal in that range could be due to reduced sensor sensitivity, rather than a true absence of interaction. 
Conversely, in the low-to-mid frequency range, a larger parameter space could potentially be ruled out, but may not be, due to underestimation of the sensor’s response.

\begin{figure}[t]
\includegraphics[width=0.9\linewidth]{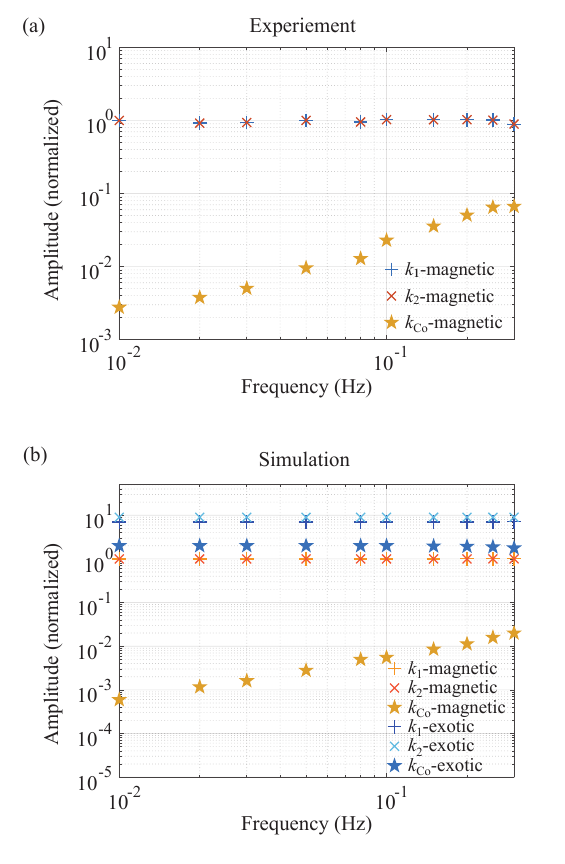}
\caption{\label{figure3} Frequency responses of the FID magnetometer and the comagnetometer. (a) Experimental variation of the output-to-input signal amplitude ratio of the FID magnetometer $k_{1}$, $k_{2}$ and comagnetometer $k_{\text{Co}}$ as a function of the oscillating magnetic field frequency. (b) Simulated variation of the output-to-input signal amplitude ratio of the FID magnetometer and comagnetometer as a function of the frequency of either an oscillating magnetic field or an exotic field–nucleon coupling.}
\end{figure}

\section{Frequency response of an FID comagnetometer}
For a single-species comagnetometer operated in open-loop state, the magnetic resonance signals corresponding to the two hyperfine levels are not obtained simultaneously, preventing simultaneous measurement of their frequency responses to external magnetic fields at the same time~\cite{wang2020single}.
A phase-lock loop can be used to independently close the loop for each energy level~\cite{wang2023atomic}.
Another approach is to use an FID comagnetometer scheme, which allows simultaneous acquisition of the precession frequencies of the two hyperfine levels through the FID signal~\cite{yang2021all,zhao2024suppression}.

In this section, we present experimental measurements of the response amplitude of the FID magnetometer and the FID comagnetometer to magnetic field perturbations at various frequencies.
Using the frequency response framework developed earlier, we perform simulations that are consistent with the experimental results.
For exotic field–nucleon couplings, the simulations reveal that, because the measurement bandwidth of the FID comagnetometer is much narrower than the intrinsic linewidth of the system, its response does not show a significant frequency dependence.
\subsection{Experimental response of an FID comagnetometer to magnetic field perturbations}
The FID comagnetometer obtains the magnetic field signals from the frequency of the two hyperfine levels and calculates their difference, focusing on the frequency response of the differential signal to the perturbation field.
Performing a Fourier transform on the FID signal within each period ($\sim$1.5 s in our experiment), the imaginary part is fitted using the superposition of two Lorentzian line shapes to extract the precession frequencies of the two hyperfine levels $f_1$ and $f_2$, which correspond to two magnetic field values $B_1$ and $B_2$.
By continuously acquiring data for 1000 times, we obtain two sets of magnetic field signals and the difference. 
The experimental parameters are consistent with those of our previously optimized FID comagnetometer setup~\cite{zhao2024suppression,yang2021all}. 
The amplitude of the perturbation field, 0.025 nT, is determined from calibration results.
The variation of the output-to-input signal amplitude ratio of the FID magnetometer $k_{1}$, $k_{2}$ and comagnetometer $k_{\text{Co}}$ is calculated for oscillating magnetic fields of different frequencies, with the results shown in Fig.~\ref{figure3}(a). 
Due to the sampling rate limitation, the Nyquist bandwidth is smaller than the magnetometer’s bandwidth. 
Therefore, within the Nyquist bandwidth, the response amplitude of the magnetometer to magnetic perturbations remains essentially constant.
However, since the magnetic field signal measured by the FID is an averaging process over the duration of the FID signal, higher signal frequencies result in greater magnetic field variations within one FID period. 
As a result, the amplitude measured by the comagnetometer, formed by the two hyperfine energy levels, shows a slight upward trend with increasing signal frequency, although it remains much lower than the output of a single magnetometer.

\subsection{Response simulation of an FID comagnetometer to magnetic and exotic spin-dependent perturbations}
The frequency response of the FID process is difficult to obtain analytically, so we establish simulations to analyze the responses of the FID magnetometer and the comagnetometer to magnetic and exotic field perturbations.
In an FID magnetometer, the presence of an oscillating field causes variations in the Larmor precession frequency. The signal can be described by a modified FID model (assuming a flat frequency response)~\cite{hunter2018waveform},
\begin{equation}
    P_x(t)=A e^{-R_{e} t} \sin \left[\omega_0 t+\phi_0+\gamma \int_0^t B_s(\tau) d \tau\right]. \label{eq:FID}
\end{equation}
Taking into account the frequency response and substituting
$k(\omega_{s})$ from Eq.~\eqref{eq:ki} and the exotic field $\mathbf{B_s}$ from Eq.~\eqref{eq:perturbation}, Eq.~\eqref{eq:FID} can be rewritten as

\begin{equation}
    P_x(t)=A e^{-R_{e}t} \sin \left[\omega_0 t+\phi_0+\beta \sin \left(\omega_s t+\phi_s\right)\right],
\end{equation}
where $\beta=\gamma k(\omega_{s})B_{s}/\omega_{s}$ is the modulation index, $\phi_{0}$ is the initial phase, and $\phi_{s}$ relates to the initial phase of the perturbation field.
By performing a Fourier transform on $P_{x}(t)$, the corresponding precession frequencies can be obtained, which can then be used to derive the magnetic field. 
Through continuous measurements, the time-domain signal of the magnetic field can be acquired, allowing us to obtain the FID magnetometer's response to magnetic field perturbations.

For a comagnetometer operated in an FID scheme, the atomic polarization can be expressed as
\begin{equation}
\begin{aligned}
    P_{x}(t)=&A_{1}e^{-R_{e,1}t}\sin[\omega_{1}t+\phi_{1}+\beta_{1}\sin(\omega_{s}t+\phi_{s})]\\
    +&A_{2}e^{-R_{e,2}t}\sin[\omega_{2}t+\phi_{2}+\beta_{2}\sin(\omega_{s}t+\phi_{s})].
\end{aligned}\label{eq:FIDP}
\end{equation}
For a single-species comagnetometer, the subscripts $1$, $2$ represent the two hyperfine levels.
We simulate the FID signal of the comagnetometer under magnetic field perturbation based on Eq.~\eqref{eq:FIDP}, where $\phi_{s}$ is determined by the phase of $B_{s}(t)$ at the starting point of a single measurement.
The simulation parameters are shown in Table~\ref{tab:table2}.
The simulated relaxation rates are approximate values based on experimental parameters.

To simulate the responses of the FID magnetometer and the comagnetometer to exotic spin-dependent interactions, we refer to Table~\ref{tab:table1}, which shows that the response of the magnetometer to exotic fields is proportional to its response to magnetic fields. 
Taking the exotic field-nucleon coupling as an example, and considering the contributions from $K_a=2I\cdot\gamma/\gamma_{a}$ and $K_b=(2I+2)\cdot\gamma/\gamma_{b}$ to $\beta_{1}$ and $\beta_2$,
we obtain the amplitude response of the FID comagnetometer, as shown in Fig.~\ref{figure3}(b).
It presents the simulated frequency responses of the FID magnetometer and the comagnetometer to magnetic perturbations (yellow) and exotic field-nucleon couplings (blue). 
The simulation results for magnetic perturbations align closely with experimental trends. 
For exotic field perturbations, it can be seen that the amplitude responses of both the FID magnetometer and the FID comagnetometer remain constant with respect to the perturbation field frequency, indicating that the frequency of the exotic field does not need to be considered when converting the measurement results to the coupling strength.

\begin{table}[t]
\renewcommand{\arraystretch}{1.3}
\caption{\label{tab:table2}%
 The complete list of parameters used for the simulations.
}
\begin{ruledtabular}
\begin{tabular}{cc}
\textrm{Parameters}&
\textrm{Values}\\
\colrule
Interval of a single field measurement (s) & 1.5\\
Sampling rate for FID signals (Hz) & $5\times10^{4}$ \\
Duration of the FID signal (s) & 1\\
Bias field magnitude (nT) & 5000\\
Oscillating field magnitude (nT)&0.025\\
Relaxation rate $R_{e,1}$ (Hz)& $2\pi\times1.5 $\\
Relaxation rate $R_{e,2}$ (Hz)& $2\pi\times 2.5$\\
Gyromagnetic ratio $\gamma_{1}$ (Hz/nT)& $2\pi\times3.4986$\\
Gyromagnetic ratio $\gamma_{2}$ (Hz/nT)& $2\pi\times3.5098$\\
\end{tabular}
\end{ruledtabular}
\end{table}

The experimental and simulation results indicate that when the sampling rate of the comagnetometer is much lower than the intrinsic linewidth, the frequency response can be neglected in the detection of exotic interactions. 
However, if the sampling rate of the FID comagnetometer is increased or the linewidth of the comagnetometer is reduced, the influence of frequency response must be taken into account, as discussed in Section~\ref{se2}.
For a detailed explanation, please refer to Appendix~\ref{app:Simu}.

\section{Calibration protocol}
Although detecting signals related to dark matter can reveal the presence of anomalous fields, the amplitude of the anomalous field cannot be determined because the coupling constant between exotic fields and subatomic particles is unknown. 
Calibration for exotic field searches is challenging because generating a known exotic field source is typically not feasible.
For magnetometers, it is possible to send pulses of known amplitude and compare the measurement results with known values.
However, the calibration method can only test the system's response to magnetic signals. 
To simulate non-magnetic signals, BGO crystals ($\text{Bi}_4\text{Ge}_3\text{O}_{12}$) are commonly used as unpolarized nucleon
sources for searching for exotic spin-dependent interactions~\cite{xiao2023femtotesla}.
Rotation and light shifts can also be used as calibration fields~\cite{padniuk2024universal,rosenzweig2024atomic}.

To simulate non-magnetic signals, an equivalent virtual magnetic field generated by vector light shift can be used. 
The equivalent virtual magnetic field interacts with electrons but does not interact with protons and neutrons, i.e., $\mathbf{b_{e}}=\mathbf{L}$, $\mathbf{b_{n}}=0$, and $\mathbf{b_{p}}=0$.
For circularly polarized light, the expression for the equivalent virtual magnetic field corresponding to the vector light shift is
\begin{equation}
    \mathbf{L}=\frac{\delta E}{h\gamma}\hat{z}=-\frac{1}{h\gamma}\frac{\alpha_{F}(\omega)}{2F}\left(\frac{E_{0}}{2}\right)^2\hat{z},
\end{equation}
where $\delta E$ represents the energy shift caused by light shift, $E_{0}$ is the light field intensity, related to the light power, and $\alpha_{F}(\omega)$ is the dynamic polarization, which depends on the frequency and other factors~\cite{cohen1972experimental,zhao2024suppression}.
Therefore, under constant conditions, a circularly polarized light beam can be used, with its amplitude modulated by an acousto-optic modulator (AOM) at a certain frequency $\omega_{s}$, thereby simulating an anomalous field-electron interaction with an oscillation frequency of $\omega_{s}$.

Although the comagnetometer’s response to anomalous field-electron coupling is similar to its response to magnetic field perturbations, as shown in Table~\ref{tab:table1}, the light shift affects the two hyperfine levels of the comagnetometer differently. 
Therefore, the response of the comagnetometer to the equivalent virtual magnetic field generated by the light shift differs from its response to the magnetic field, allowing the light shift to be used as a calibration field. 
For the cesium $D_1$ line, the light shifts of the two ground-state hyperfine levels are calculated in detail in the appendix of reference~\cite{zhao2024suppression}.
If the laser frequency is tuned to be resonant with the $F_{g}=4\rightarrow F_{e}=3$ transition of the $D_1$ line, the light shift on the $F_{g}=3$ level becomes negligible and can be safely ignored, i.e., $\mathbf{b_{e,3}}\approx0$, allowing as to consider only the light shift of the $F_{g}=4$ level,
\begin{equation}
    \mathbf{b_{e,4}}=b_{e,4}\cos(\omega_s t)\hat{z}=-\frac{\alpha_{4}(\omega)}{8h\gamma_{a}}\left(\frac{E_{0}}{2}\right)^2\cos(\omega_s t)\hat{z}.
\end{equation}
Therefore, based on the theoretical frequency response described above, the amplitude of the differential signal of the comagnetometer should be given by
\begin{equation}
    S_{\text{Co}}(\omega_s)=\frac{\gamma}{\gamma_{a}}\frac{R_{e,a}}{\sqrt{\omega_{s}^{2}+R_{e,a}^{2}}}b_{e,4}.
\end{equation}
By varying the modulation frequency of the light intensity, the theoretical prediction can be compared with the experimentally measured signal amplitude of the comagnetometer, thereby verifying its frequency response to non-magnetic forces.

\section{Discussion and Outlook}
In this work, we present a theoretical and experimental investigation of the frequency response of a differential comagnetometer to oscillating magnetic signals and exotic spin-dependent couplings.
We analyze the relationship between the comagnetometer’s frequency response to magnetic and non-magnetic signals. 
Based on the experimentally measured magnetic response, we outline a method to determine the comagnetometer’s response to non-magnetic interactions and propose a procedure to infer the strength of exotic interactions from a comagnetometer, taking the axion wind as a representative example. 
The frequency response analysis presented here allows for a revision of the limits on exotic field coupling strength previously obtained using comagnetometers.
In addition, we introduce an experimental protocol that uses light shifts as calibration pulses to verify the comagnetometer’s frequency response.
To the best of our knowledge, the frequency response of differential comagnetometers to non-magnetic signals has not yet been systematically explored. 
Our method provides a crucial basis for accurately evaluating and improving the sensitivity of comagnetometers to exotic spin-dependent interactions.

The frequency response analysis method proposed in this work is broadly applicable to various sensors designed to probe exotic spin-dependent interactions, with comagnetometers serving as a representative example.
In addition, it provides insights into the analysis of gradiometers, facilitating the differential detection of biological magnetic signals~\cite{zhang2020recording, wu2024effects}.
By precisely controlling the pulse sequence, the bandwidth of the FID magnetometer can be enhanced to the kilohertz level~\cite{li2022kilohertz,yi2024free}.
If the FID comagnetometer achieves such linewidth, it will remain insensitive to magnetic field changes over a wider bandwidth range while maintaining sensitivity to exotic field perturbations.
Due to its significant difference in frequency response compared to self-compensating comagnetometers, the two types of comagnetometers can complement each other in network detection for ultralight bosonic dark matter~\cite{afach2021search,afach2024can}, thus playing a unique role at high bandwidths and helping to identify the coupling type.

\begin{acknowledgments}
We thankfully acknowledge useful discussions with Yufei Zhang.
This work is supported by the National Natural Science Foundation of China (No. 62071012), the National Science Fund for Distinguished Young Scholars of China (Grant No. 61225003), and National Hi-Tech Research and Development (863) Program. T. W. acknowledges the support from the start-up funding for young researchers of Peking University.
\end{acknowledgments}

\appendix

\section{The evolution equation for the electron polarization}\label{app:dP}

In the interaction picture, considering the effects of the applied magnetic field $\mathbf{B}$ and the exotic spin-related field $\mathbf{\Xi}$, the evolution of the electron spin $\mathbf{S}$ and the nuclear spin $\mathbf{I}$ of the ground-state alkali metal can be described by the interaction Hamiltonian $H$~\cite{padniuk2022response},
\begin{equation}
  H=\mu_B g_S \mathbf{B} \cdot \mathbf{S}+\mu_B g_I \mathbf{B} \cdot \mathbf{I}+\chi_e \boldsymbol{\Xi} \cdot \mathbf{S}+\chi_I \boldsymbol{\Xi} \cdot \mathbf{I},  \label{eq:H1}
\end{equation}
where $\mu_B$ is the Bohr magneton, $g_s$ and $g_I$ are the electron and nuclear Landé factors, respectively, and $\chi_{e}$ and $\chi_{I}$ are the coupling constants of the electron and nuclear to the exotic field.
Decompose the coupling constant $\chi_{I}$ into proton and neutron contributions
\begin{equation}
    \chi_I=\sigma_p \chi_p+\sigma_n \chi_n,\label{eq:chi_I}
\end{equation}
where $\sigma_{p}$ and $\sigma_{n}$ represent the contributions of the proton and neutron spins to the overall nuclear spin of the alkali metal, while $\chi_{p}$ and $\chi_{n}$ are the coupling constants for the proton and neutron with the exotic field, respectively. 
To facilitate the comparison of the atom's spin response to magnetic field and exotic field, we redefine an equivalent pseudo-magnetic field,
\begin{subequations}
    \begin{align}
\mathbf{b}_{\mathbf{e}}=\frac{\chi_e \boldsymbol{\Xi}}{g_S \mu_B}, \\
\mathbf{b}_{\mathbf{p}}=\frac{\sigma_p \chi_p \boldsymbol{\Xi}}{g_S \mu_B},\\
\mathbf{b}_{\mathbf{n}}=\frac{\sigma_n \chi_n \boldsymbol{\Xi}}{g_S \mu_B} .
\end{align}\label{eq:psedufield}
\end{subequations}

Substituting Eqs.~([\ref{eq:chi_I}, \ref{eq:psedufield}]) in the theoretical section into Eq.~\eqref{eq:H1}, the interaction Hamiltonian can be rewritten as follows
\begin{equation}
    H=\mu_B g_S\left(\mathbf{B}+\mathbf{b}_{\mathbf{e}}\right) \cdot \mathbf{S}+\mu_B g_S\left(\frac{g_I}{g_S}\mathbf{B}+\mathbf{b}_{\mathbf{n}}+\mathbf{b}_{\mathbf{p}}\right) \cdot \mathbf{I} .
\end{equation}
From the above equation, it can be seen that the interaction between alkali metal atoms and both the magnetic field or the exotic field can be decomposed into two parts: the interaction of the electron spin and the interaction of the nuclear spin.

The Heisenberg equation of motion for the observable quantity is
\begin{equation}
    \frac{\mathrm{d} \mathcal{A}}{\mathrm{d} t}=-\frac{1}{i \hbar}[H,\mathcal{A}],
\end{equation}
and the time evolution equation of the total angular momentum $F$ of the alkali metal atom can be written as
\begin{equation}
\begin{aligned}
\frac{\mathrm{d} \mathbf{F}}{\mathrm{d} t} & =-\frac{\mu_B g_S}{i \hbar}\left[\left(\mathbf{B}+\mathbf{b}_{\mathbf{e}}\right) \cdot \mathbf{S}+\left(\frac{g_I}{g_S}\mathbf{B}+\mathbf{b}_{\mathbf{n}}+\mathbf{b}_{\mathbf{p}}\right) \cdot \mathbf{I}, \mathbf{F}\right] \\
& =\gamma_e\left(\mathbf{B}+\mathbf{b}_{\mathbf{e}}\right) \times \mathbf{S}+\gamma_e\left(\frac{g_I}{g_S}\mathbf{B}+\mathbf{b}_{\mathbf{n}}+\mathbf{b}_{\mathbf{p}}\right) \times \mathbf{I}.
\end{aligned}
\end{equation}
The time evolution of the expectation value of angular momentum
\begin{equation}
    \frac{\mathrm{d}\langle\mathbf{F}\rangle}{\mathrm{d} t}=\gamma_e\left(\mathbf{B}+\mathbf{b}_{\mathbf{e}}\right) \times\langle\mathbf{S}\rangle+\gamma_e\left(\frac{g_I}{g_S}\mathbf{B}+\mathbf{b}_{\mathbf{n}} +\mathbf{b}_{\mathbf{p}} \right) \times\langle\mathbf{I}\rangle .\label{eq:dF1}
\end{equation}
According to the projection theorem,
\begin{equation}
\begin{aligned}
    \langle\mathbf{S}\rangle & =\frac{\langle\mathbf{S} \cdot \mathbf{F}\rangle}{F(F+1)}\langle\mathbf{F}\rangle \\
    & =\frac{\left\langle\mathbf{S}^2+\mathbf{F}^2-\mathbf{I}^2\right\rangle}{2 F(F+1)}\langle\mathbf{F}\rangle,\label{eq:ProThe}
\end{aligned}
\end{equation}
for the ground-state hyperfine levels of alkali metals $F_{a,b}$, 
\begin{equation}
    \langle\mathbf{S}_{a,b}\rangle=\pm\frac{1}{2I+1}\langle\mathbf{F}_{a,b}\rangle.\label{eq:Sab}
\end{equation}
Since $\langle\mathbf{S}_{a,b}\rangle+\langle\mathbf{I}_{a,b}\rangle=\langle\mathbf{F}_{a,b}\rangle$,
\begin{equation}
    \langle\mathbf{I}_{a,b}\rangle=\left(1\mp\frac{1}{2I+1}\right)\langle\mathbf{F}_{a,b}\rangle.\label{eq:Iab}
\end{equation}
Substitute Eqs.~([\ref{eq:Sab}, \ref{eq:Iab}]) into Eq.~\eqref{eq:dF1}, and replace the expectation value of the total angular momentum with the expectation value of the electron spin angular momentum, we can derive
\begin{equation}
\begin{aligned}
    \frac{\mathrm{d}\left\langle\mathbf{S}_{\mathbf{a}, \mathbf{b}}\right\rangle}{\mathrm{d} t}&= \pm \frac{\gamma_e}{2 I+1}\left(\mathbf{B}+\mathbf{b}_{\mathbf{e}}\right) \times\left\langle\mathbf{S}_{\mathbf{a}, \mathbf{b}}\right\rangle\\
    &+\gamma_e\left(1 \mp \frac{1}{2 I+1}\right)\left(\frac{g_I}{g_S}\mathbf{B}+\mathbf{b}_{\mathbf{n}}+\mathbf{b}_{\mathbf{p}}\right) \times\left\langle\mathbf{S}_{\mathbf{a}, \mathbf{b}}\right\rangle.
\end{aligned}
\end{equation}
Introduce the electron’s polarizability
\begin{equation}
    \mathbf{P}=\frac{\left\langle\mathbf{S}_{\mathbf{a}, \mathbf{b}}\right\rangle}{S},
\end{equation}
and consider the relaxation rate $R_{e}$ and the pumping rate $R_{p}$
of electron polarization, the evolution equation of electron polarization in an all-optical magnetometer can be written as Eq.~\eqref{eq:dP1}.

\section{Perturbative solution of the polarization evolution equation}\label{app:perturbation}
Expanding Eq.~\eqref{eq:dPv} into the form of components along the three directions, and set $B_{z}(t)=B_{0}+B_{s}\cos(\omega_{s}t)$, we derive
\begin{subequations}
    \begin{align}
        \frac{\mathrm{d} P_x}{\mathrm{d} t} &=-\gamma_a B_{z}(t) P_y-(R_e+P_{p}(t)) P_x+R_p(t)P_{0},\\
        \frac{\mathrm{d} P_y}{\mathrm{d} t} &=\gamma_a B_{z}(t) P_x-(R_e+P_{p}(t)) P_y,\\
        \frac{\mathrm{d} P_z}{\mathrm{d} t} &=-(R_e+P_{p}(t)) P_z .
    \end{align}
\end{subequations}

To simplify the calculation and obtain an analytical solution, we rewrite the perturbation field as $B_{s}e^{i\omega_{s} t}$, and rewrite the optical pumping rate as $R_{p}e^{i\omega t}$.
Expand $\mathbf{P}$ into the components along the three directions, let $P_{+}=P_{x}+iP_{y}$, and transfer it to the rotating frame as $ \tilde{P}_{+}=P_{+}e^{-i\omega t}$.
Using the rotating wave approximation, we can neglect the rapidly oscillating terms and ultimately obtain the evolution of $\tilde{P}_{+}$ 

\begin{equation}
    \frac{\mathrm{d} \tilde{P}_{+}}{\mathrm{d} t}=i\left[\gamma_{a}(B_{0}+B_s e^{i\omega_{s} t})-\omega\right]\tilde{P}_{+}-R_{e}\tilde{P}_{+}+R_{p}P_{0}.\label{eq:dP2}
\end{equation}
The problem of solving the polarization response to the perturbation field is simplified to a first-order non-homogeneous linear differential equation.
Expand $\tilde{P}_{+}$ perturbatively in terms of $\gamma_{a}B_{s}/{R_{e}}$,
\begin{equation}
    \tilde{P}_{+}=\sum_{n=0}^{+\infty}\tilde{P}_{+}^{(n)}.\label{eq:expansion}
\end{equation}
We can then obtain the evolution equations for $\text{d}P_{+}$ and $\mathrm{d} \tilde{P}_{+}$ as Eq.~\eqref{eq:dP2}. Substitute the perturbative expansion Eq.~\eqref{eq:expansion} into the evolution equation Eq.~\eqref{eq:dP2}, and organize it according to the order of $\gamma_{a} B_{s}/R_{e}$,
\begin{subequations}
    \begin{align}
            \frac{\mathrm{d} \tilde{P}_{+}^{(0)}}{\mathrm{d} t}&=\left(i \gamma_{a}B_0-i\omega-R_e\right) \tilde{P}_{+}^{(0)}+R_pP_{0},\label{eq:dP0P}\\
            \frac{\mathrm{~d} \tilde{P}_{+}^{(1)}}{\mathrm{d} t}&=\left(i \gamma_{a}B_0-i\omega-R_e\right) \tilde{P}_{+}^{(1)}+i \gamma_{a} B_s e^{i \omega_s t} \tilde{P}_{+}^{(0)}.\label{eq:dP1P}
    \end{align}
\end{subequations}
Solve the differential equations sequentially for $\tilde{P}_{+}^{(0)}$ and $\tilde{P}_{+}^{(1)}$. For $\tilde{P}_{+}^{(0)}$, retain the stead-state solution,
\begin{equation}
    \tilde{P}_{+}^{(0)} = \frac{R_p P_{0}}{i\left(\omega-\gamma_{a}B_0\right)+R_e}.\label{eq:tP0}
\end{equation}
Thus, the expression for ${P}_{+}^{(0)}$ is 
\begin{equation}
    P_{+}^{(0)} = \tilde{P}_{+}^{(0)}e^{i\omega t}=\frac{R_p P_{0}e^{i\omega t}}{i\left(\omega-\gamma_{a}B_0\right)+R_e},\label{eq:P0}
\end{equation}
corresponding to the electron polarization in the absence of perturbation. 
Substitute Eq.~\eqref{eq:tP0} into Eq.~\eqref{eq:dP1P}, 
\begin{equation}
    \tilde{P}_{+}^{(1)}=\frac{i \gamma_a B_s R_pP_{0}e^{i\omega_{s} t}}{\left[i\left(\omega+\omega_s-\gamma_a B_0\right)+R_e\right]\left[i\left(\omega-\gamma_a B_0\right)+R_e\right]}.
\end{equation}
And then, we can derive the expression for $P_{+}^{(1)}$,
\begin{equation}
\begin{aligned}
     {P}_{+}^{(1)} &= \tilde{P}_{+}^{(1)}e^{i\omega t}\\
     &=\frac{i \gamma_a B_s R_pP_{0}e^{i(\omega_{s}+\omega) t}}{\left[i\left(\omega+\omega_s-\gamma_a B_0\right)+R_e\right]\left[i\left(\omega-\gamma_a B_0\right)+R_e\right]}.\label{eq:P1}
\end{aligned}
\end{equation}

Considering $P_{x}=\operatorname{Re}(P_{+})$, $P_{y}=\operatorname{Im}(P_{+})$, we can obtain the frequency response of atomic polarization to the magnetic field perturbation.
Using $\omega$ as a reference frequency to demodulate the signal $P_{x}^{(0)}$, the real and imaginary parts of the demodulated signal will correspond to the Lorentzian absorption and dispersion lines, respectively, as a function of the modulation frequency $\omega$.
When $\omega=\gamma_{a}B_{0}$, the demodulated result of $P_{x}^{(0)}$ correspond to a static signal, while for $P_{x}^{(1)}$, the  quadrature component of the demodulated signal can be expressed as
\begin{equation}
    \begin{aligned}
        S_{\text{Qu},a}^{(1)}&=\frac{\omega_{s}\cos(\omega_{s}t)-R_{e}\sin(\omega_{s}t)}{(\omega_{s}^{2}+R_{e}^{2})R_{e}}\cdot\gamma_{a}B_s R_p P_{0}\\
        &=\frac{\gamma_{a} B_s R_pP_{0}}{R_e \sqrt{\omega_s^2+R_e^2}}\sin(\omega_{s}t+\varphi),
    \end{aligned}
\end{equation}
where $\varphi$ represents the initial phase, satisfying
\begin{subequations}
    \begin{align}
        \cos \varphi&=\frac{-R_e}{\sqrt{\omega_s^2+R_e^2}}, \\
         \sin \varphi&=\frac{\omega_s}{\sqrt{\omega_s^2+R_e^2}}.
    \end{align}
\end{subequations}

\section{Experiment on measuring the response of 
a magnetometer}\label{app:exp}
To measure the response of a magnetometer, we use a Helmholtz coil to generate a small time-varying magnetic field along the bias magnetic field direction of the magnetometer.
By varying the perturbation field frequency, we measure the system's response to the perturbation field with different frequencies.

In the experiment, we measure the open loop frequency response of the Cs magnetometer. 
The paraffin-coated Cs vapor cell is placed inside a magnetic shield.
We use an amplitude-modulated Bell-Bloom magnetometer, and employ circularly polarized pump light resonant with the Cs $D_{1}$ line and linearly polarized probe light detuned from the Cs $D_{2}$ line. 
The intensity of the pump light is modulated, and when the modulation frequency matches the Larmor precession frequency, the amplitude of the detected polarization rotation signal of the linearly polarized probe light reaches its maximum.
The optical rotation signal of the probe beam is fed into a lock-in amplifier (Stanford Research Systems, SR865A), where the pump light modulation frequency is scanned around the Larmor precession frequency.
The modulation frequency is used as the reference to demodulate the optical rotation signal, yielding a Lorentzian absorption profile (X component) and a dispersion profile (Y component).
The bias magnetic field is set to approximately 5800 nT to ensure that the two signals corresponding to different hyperfine levels are fully separated while minimizing the influence of the nonlinear Zeeman effect.

We measure the frequency response of the magnetometer constructed from a single hyperfine level.
For $F_{a}=4$, fix the pump light modulation frequency at $\omega=\gamma_{a}B_{0}$.
Apply an oscillating magnetic field using Helmholtz coils, with $B_{s}$ set to approximately $0.025 \ \text{nT}$ to ensure that the variation in the Larmor frequency remains within the resonance region of the magnetic resonance signal.
The output of the Y channel of the lock-in amplifier is collected, as shown in Fig.~\ref{figure1}(b).
The voltage signal can be converted into a frequency signal through the slope of the Lorentzian dispersion line shape, which is then further converted into a magnetic field signal.
By varying the frequency of the oscillating field, different output signals are obtained.
As shown in Fig.~\ref{figure1}(b), the higher the frequency of the oscillating signal, the smaller the signal amplitude, indicating that the magnetometer's ability to reconstruction of the perturbation field is constrained by its intrinsic bandwidth.

We vary the oscillating field frequency from 0.1 Hz-15 Hz, fit the output signal using a sine function, record the signal amplitude.
Convert voltage signals to magnetic field signals and normalize them with the input signal amplitude.
The relationship between normalized amplitude and perturbation field frequency is shown in Fig.~\ref{figure1}(c).

\section{Response simulation of an FID comagnetometer to exotic couplings}\label{app:Simu}

To illustrate that when the signal frequency is much smaller than the comagnetometer linewidth, the response amplitude of the FID comagnetometer to an exotic field perturbation can be regarded as approximately constant, we perform simulations to compute the normalized output amplitude of the comagnetometer at 0.02 Hz under different relaxation rates.
To control variables, we fix the ratio $R_{e,1}:R_{e,2}$ = 3:5. The simulation results are summarized in Table~\ref{tab:table3}. 
The normalized response refers to the ratio of the output to input signal amplitudes of the comagnetometer, which ideally equals 2 (see Section~\ref{se2} B for details).
Here, difference is calculated as the percentage ratio between the relative deviation and the ideal value.
Since the simulation serves as a representation of the experimental process, individual results may differ slightly between runs, but the overall trend remains stable.

Based on Table~\ref{tab:table3}, it can be seen that when the signal frequency is much smaller than the comagnetometer linewidth, the response amplitude of the FID comagnetometer to exotic field perturbations can be regarded as approximately constant. 
However, when the signal frequency is comparable to the linewidth, the frequency dependence of the response amplitude must be taken into account. 
In other words, even if the sampling rate of the comagnetometer is increased, for high-frequency signals, it remains necessary to apply the conclusions of this work and consider the variation of the response amplitude with frequency.

\begin{table}[t]
\renewcommand{\arraystretch}{1.3}
\caption{\label{tab:table3}%
 Simulated normalized amplitude response of an FID comagnetometer to exotic couplings with different relaxation rates.
}
\begin{ruledtabular}
\begin{tabular}{cccc}
\textrm{$R_{e,1}$(Hz)}&
\textrm{$R_{e,2}$(Hz)}&
\textrm{Normalized response}&\textrm{Difference(\%)}
\\
\colrule
2$\pi\times$1.5&2$\pi\times$2.5&1.976&1.2\\
2$\pi\times$0.6&2$\pi\times$1&2.024&1.2\\
2$\pi\times$0.3&2$\pi\times$0.5&2.112&5.6\\
2$\pi\times$0.15&2$\pi\times$0.25&2.168&8.4\\
\end{tabular}
\end{ruledtabular}
\end{table}



\begin{thebibliography}{53}%
\makeatletter
\providecommand \@ifxundefined [1]{%
 \@ifx{#1\undefined}
}%
\providecommand \@ifnum [1]{%
 \ifnum #1\expandafter \@firstoftwo
 \else \expandafter \@secondoftwo
 \fi
}%
\providecommand \@ifx [1]{%
 \ifx #1\expandafter \@firstoftwo
 \else \expandafter \@secondoftwo
 \fi
}%
\providecommand \natexlab [1]{#1}%
\providecommand \enquote  [1]{``#1''}%
\providecommand \bibnamefont  [1]{#1}%
\providecommand \bibfnamefont [1]{#1}%
\providecommand \citenamefont [1]{#1}%
\providecommand \href@noop [0]{\@secondoftwo}%
\providecommand \href [0]{\begingroup \@sanitize@url \@href}%
\providecommand \@href[1]{\@@startlink{#1}\@@href}%
\providecommand \@@href[1]{\endgroup#1\@@endlink}%
\providecommand \@sanitize@url [0]{\catcode `\\12\catcode `\$12\catcode `\&12\catcode `\#12\catcode `\^12\catcode `\_12\catcode `\%12\relax}%
\providecommand \@@startlink[1]{}%
\providecommand \@@endlink[0]{}%
\providecommand \url  [0]{\begingroup\@sanitize@url \@url }%
\providecommand \@url [1]{\endgroup\@href {#1}{\urlprefix }}%
\providecommand \urlprefix  [0]{URL }%
\providecommand \Eprint [0]{\href }%
\providecommand \doibase [0]{https://doi.org/}%
\providecommand \selectlanguage [0]{\@gobble}%
\providecommand \bibinfo  [0]{\@secondoftwo}%
\providecommand \bibfield  [0]{\@secondoftwo}%
\providecommand \translation [1]{[#1]}%
\providecommand \BibitemOpen [0]{}%
\providecommand \bibitemStop [0]{}%
\providecommand \bibitemNoStop [0]{.\EOS\space}%
\providecommand \EOS [0]{\spacefactor3000\relax}%
\providecommand \BibitemShut  [1]{\csname bibitem#1\endcsname}%
\let\auto@bib@innerbib\@empty
\bibitem [{\citenamefont {Ripka}(2002)}]{ripka2002magnetic}%
  \BibitemOpen
  \bibfield  {author} {\bibinfo {author} {\bibfnamefont {P.}~\bibnamefont {Ripka}},\ }\bibfield  {title} {\bibinfo {title} {Magnetic sensors and magnetometers},\ }\href {https://doi.org/10.1088/0957-0233/13/4/707} {\bibfield  {journal} {\bibinfo  {journal} {Meas. Sci. Technol}\ }\textbf {\bibinfo {volume} {13}},\ \bibinfo {pages} {645} (\bibinfo {year} {2002})}\BibitemShut {NoStop}%
\bibitem [{\citenamefont {Bertoldi}\ \emph {et~al.}(2005)\citenamefont {Bertoldi}, \citenamefont {Bassi}, \citenamefont {Ricci}, \citenamefont {Covi},\ and\ \citenamefont {Varas}}]{bertoldi2005magnetoresistive}%
  \BibitemOpen
  \bibfield  {author} {\bibinfo {author} {\bibfnamefont {A.}~\bibnamefont {Bertoldi}}, \bibinfo {author} {\bibfnamefont {D.}~\bibnamefont {Bassi}}, \bibinfo {author} {\bibfnamefont {L.}~\bibnamefont {Ricci}}, \bibinfo {author} {\bibfnamefont {D.}~\bibnamefont {Covi}},\ and\ \bibinfo {author} {\bibfnamefont {S.}~\bibnamefont {Varas}},\ }\bibfield  {title} {\bibinfo {title} {Magnetoresistive magnetometer with improved bandwidth and response characteristics},\ }\href {https://doi.org/10.1063/1.1922787} {\bibfield  {journal} {\bibinfo  {journal} {Rev. Sci. Instrum.}\ }\textbf {\bibinfo {volume} {76}},\ \bibinfo {pages} {065106} (\bibinfo {year} {2005})}\BibitemShut {NoStop}%
\bibitem [{\citenamefont {Bevilacqua}\ \emph {et~al.}(2016)\citenamefont {Bevilacqua}, \citenamefont {Biancalana}, \citenamefont {Chessa},\ and\ \citenamefont {Dancheva}}]{bevilacqua2016multichannel}%
  \BibitemOpen
  \bibfield  {author} {\bibinfo {author} {\bibfnamefont {G.}~\bibnamefont {Bevilacqua}}, \bibinfo {author} {\bibfnamefont {V.}~\bibnamefont {Biancalana}}, \bibinfo {author} {\bibfnamefont {P.}~\bibnamefont {Chessa}},\ and\ \bibinfo {author} {\bibfnamefont {Y.}~\bibnamefont {Dancheva}},\ }\bibfield  {title} {\bibinfo {title} {Multichannel optical atomic magnetometer operating in unshielded environment},\ }\href {https://doi.org/https://doi.org/10.1007/s00340-016-6375-2} {\bibfield  {journal} {\bibinfo  {journal} {Appl. Phys. B}\ }\textbf {\bibinfo {volume} {122}},\ \bibinfo {pages} {1} (\bibinfo {year} {2016})}\BibitemShut {NoStop}%
\bibitem [{\citenamefont {Bevilacqua}\ \emph {et~al.}(2021)\citenamefont {Bevilacqua}, \citenamefont {Biancalana}, \citenamefont {Dancheva}, \citenamefont {Fregosi},\ and\ \citenamefont {Vigilante}}]{bevilacqua2021spin}%
  \BibitemOpen
  \bibfield  {author} {\bibinfo {author} {\bibfnamefont {G.}~\bibnamefont {Bevilacqua}}, \bibinfo {author} {\bibfnamefont {V.}~\bibnamefont {Biancalana}}, \bibinfo {author} {\bibfnamefont {Y.}~\bibnamefont {Dancheva}}, \bibinfo {author} {\bibfnamefont {A.}~\bibnamefont {Fregosi}},\ and\ \bibinfo {author} {\bibfnamefont {A.}~\bibnamefont {Vigilante}},\ }\bibfield  {title} {\bibinfo {title} {Spin dynamic response to a time dependent field},\ }\href {https://doi.org/https://doi.org/10.1007/s00340-021-07673-y} {\bibfield  {journal} {\bibinfo  {journal} {Appl. Phys. B}\ }\textbf {\bibinfo {volume} {127}},\ \bibinfo {pages} {128} (\bibinfo {year} {2021})}\BibitemShut {NoStop}%
\bibitem [{\citenamefont {Zhang}\ \emph {et~al.}(2020{\natexlab{a}})\citenamefont {Zhang}, \citenamefont {Wu}, \citenamefont {Chen}, \citenamefont {Peng},\ and\ \citenamefont {Guo}}]{zhang2020frequency}%
  \BibitemOpen
  \bibfield  {author} {\bibinfo {author} {\bibfnamefont {R.}~\bibnamefont {Zhang}}, \bibinfo {author} {\bibfnamefont {T.}~\bibnamefont {Wu}}, \bibinfo {author} {\bibfnamefont {J.}~\bibnamefont {Chen}}, \bibinfo {author} {\bibfnamefont {X.}~\bibnamefont {Peng}},\ and\ \bibinfo {author} {\bibfnamefont {H.}~\bibnamefont {Guo}},\ }\bibfield  {title} {\bibinfo {title} {Frequency response of optically pumped magnetometer with nonlinear \text{Zeeman} effect},\ }\href {https://doi.org/https://doi.org/10.3390/app10207031} {\bibfield  {journal} {\bibinfo  {journal} {Appl. Sci.}\ }\textbf {\bibinfo {volume} {10}},\ \bibinfo {pages} {7031} (\bibinfo {year} {2020}{\natexlab{a}})}\BibitemShut {NoStop}%
\bibitem [{\citenamefont {Li}\ \emph {et~al.}(2020)\citenamefont {Li}, \citenamefont {Baynes}, \citenamefont {Luiten},\ and\ \citenamefont {Perrella}}]{li2020continuous}%
  \BibitemOpen
  \bibfield  {author} {\bibinfo {author} {\bibfnamefont {R.}~\bibnamefont {Li}}, \bibinfo {author} {\bibfnamefont {F.~N.}\ \bibnamefont {Baynes}}, \bibinfo {author} {\bibfnamefont {A.~N.}\ \bibnamefont {Luiten}},\ and\ \bibinfo {author} {\bibfnamefont {C.}~\bibnamefont {Perrella}},\ }\bibfield  {title} {\bibinfo {title} {Continuous high-sensitivity and high-bandwidth atomic magnetometer},\ }\href {https://doi.org/https://doi.org/10.1103/PhysRevApplied.14.064067} {\bibfield  {journal} {\bibinfo  {journal} {Phys. Rev. Appl.}\ }\textbf {\bibinfo {volume} {14}},\ \bibinfo {pages} {064067} (\bibinfo {year} {2020})}\BibitemShut {NoStop}%
\bibitem [{\citenamefont {Zhang}(2025)}]{zhang2025influence}%
  \BibitemOpen
  \bibfield  {author} {\bibinfo {author} {\bibfnamefont {R.}~\bibnamefont {Zhang}},\ }\bibfield  {title} {\bibinfo {title} {Influence of atomic magnetometer’s orientation on its frequency response},\ }\href {https://doi.org/https://doi.org/10.3390/s25051364} {\bibfield  {journal} {\bibinfo  {journal} {Sensors}\ }\textbf {\bibinfo {volume} {25}},\ \bibinfo {pages} {1364} (\bibinfo {year} {2025})}\BibitemShut {NoStop}%
\bibitem [{\citenamefont {Abel}\ \emph {et~al.}(2017)\citenamefont {Abel}, \citenamefont {Ayres}, \citenamefont {Ban}, \citenamefont {Bison}, \citenamefont {Bodek}, \citenamefont {Bondar}, \citenamefont {Daum}, \citenamefont {Fairbairn}, \citenamefont {Flambaum}, \citenamefont {Geltenbort} \emph {et~al.}}]{abel2017search}%
  \BibitemOpen
  \bibfield  {author} {\bibinfo {author} {\bibfnamefont {C.}~\bibnamefont {Abel}}, \bibinfo {author} {\bibfnamefont {N.~J.}\ \bibnamefont {Ayres}}, \bibinfo {author} {\bibfnamefont {G.}~\bibnamefont {Ban}}, \bibinfo {author} {\bibfnamefont {G.}~\bibnamefont {Bison}}, \bibinfo {author} {\bibfnamefont {K.}~\bibnamefont {Bodek}}, \bibinfo {author} {\bibfnamefont {V.}~\bibnamefont {Bondar}}, \bibinfo {author} {\bibfnamefont {M.}~\bibnamefont {Daum}}, \bibinfo {author} {\bibfnamefont {M.}~\bibnamefont {Fairbairn}}, \bibinfo {author} {\bibfnamefont {V.~V.}\ \bibnamefont {Flambaum}}, \bibinfo {author} {\bibfnamefont {P.}~\bibnamefont {Geltenbort}}, \emph {et~al.},\ }\bibfield  {title} {\bibinfo {title} {Search for axionlike dark matter through nuclear spin precession in electric and magnetic fields},\ }\href {https://doi.org/https://doi.org/10.1103/PhysRevX.7.041034} {\bibfield  {journal} {\bibinfo  {journal} {Phys. Rev. X}\ }\textbf {\bibinfo {volume} {7}},\ \bibinfo {pages} {041034} (\bibinfo {year}
  {2017})}\BibitemShut {NoStop}%
\bibitem [{\citenamefont {Safronova}\ \emph {et~al.}(2018)\citenamefont {Safronova}, \citenamefont {Budker}, \citenamefont {DeMille}, \citenamefont {Kimball}, \citenamefont {Derevianko},\ and\ \citenamefont {Clark}}]{safronova2018search}%
  \BibitemOpen
  \bibfield  {author} {\bibinfo {author} {\bibfnamefont {M.}~\bibnamefont {Safronova}}, \bibinfo {author} {\bibfnamefont {D.}~\bibnamefont {Budker}}, \bibinfo {author} {\bibfnamefont {D.}~\bibnamefont {DeMille}}, \bibinfo {author} {\bibfnamefont {D.~F.~J.}\ \bibnamefont {Kimball}}, \bibinfo {author} {\bibfnamefont {A.}~\bibnamefont {Derevianko}},\ and\ \bibinfo {author} {\bibfnamefont {C.~W.}\ \bibnamefont {Clark}},\ }\bibfield  {title} {\bibinfo {title} {Search for new physics with atoms and molecules},\ }\href {https://doi.org/10.1103/RevModPhys.90.025008} {\bibfield  {journal} {\bibinfo  {journal} {Rev. Mod. Phys.}\ }\textbf {\bibinfo {volume} {90}},\ \bibinfo {pages} {025008} (\bibinfo {year} {2018})}\BibitemShut {NoStop}%
\bibitem [{\citenamefont {Jackson~Kimball}\ \emph {et~al.}(2023)\citenamefont {Jackson~Kimball}, \citenamefont {Budker}, \citenamefont {Chupp}, \citenamefont {Geraci}, \citenamefont {Kolkowitz}, \citenamefont {Singh},\ and\ \citenamefont {Sushkov}}]{jackson2023probing}%
  \BibitemOpen
  \bibfield  {author} {\bibinfo {author} {\bibfnamefont {D.~F.}\ \bibnamefont {Jackson~Kimball}}, \bibinfo {author} {\bibfnamefont {D.}~\bibnamefont {Budker}}, \bibinfo {author} {\bibfnamefont {T.~E.}\ \bibnamefont {Chupp}}, \bibinfo {author} {\bibfnamefont {A.~A.}\ \bibnamefont {Geraci}}, \bibinfo {author} {\bibfnamefont {S.}~\bibnamefont {Kolkowitz}}, \bibinfo {author} {\bibfnamefont {J.~T.}\ \bibnamefont {Singh}},\ and\ \bibinfo {author} {\bibfnamefont {A.~O.}\ \bibnamefont {Sushkov}},\ }\bibfield  {title} {\bibinfo {title} {Probing fundamental physics with spin-based quantum sensors},\ }\href {https://doi.org/10.1103/PhysRevA.108.010101} {\bibfield  {journal} {\bibinfo  {journal} {Phys. Rev. A}\ }\textbf {\bibinfo {volume} {108}},\ \bibinfo {pages} {010101} (\bibinfo {year} {2023})}\BibitemShut {NoStop}%
\bibitem [{\citenamefont {Abbott}(2017)}]{abbott2017multi}%
  \BibitemOpen
  \bibfield  {author} {\bibinfo {author} {\bibfnamefont {B.~P.}\ \bibnamefont {Abbott}},\ }\bibfield  {title} {\bibinfo {title} {Multi-messenger observations of a binary neutron star merger},\ }\href {https://doi.org/10.3847/2041-8213/aa91c9} {\bibfield  {journal} {\bibinfo  {journal} {Astrophys. J. Lett.}\ }\textbf {\bibinfo {volume} {848}},\ \bibinfo {pages} {L12} (\bibinfo {year} {2017})}\BibitemShut {NoStop}%
\bibitem [{\citenamefont {Dailey}\ \emph {et~al.}(2021)\citenamefont {Dailey}, \citenamefont {Bradley}, \citenamefont {Jackson~Kimball}, \citenamefont {Sulai}, \citenamefont {Pustelny}, \citenamefont {Wickenbrock},\ and\ \citenamefont {Derevianko}}]{dailey2021quantum}%
  \BibitemOpen
  \bibfield  {author} {\bibinfo {author} {\bibfnamefont {C.}~\bibnamefont {Dailey}}, \bibinfo {author} {\bibfnamefont {C.}~\bibnamefont {Bradley}}, \bibinfo {author} {\bibfnamefont {D.~F.}\ \bibnamefont {Jackson~Kimball}}, \bibinfo {author} {\bibfnamefont {I.~A.}\ \bibnamefont {Sulai}}, \bibinfo {author} {\bibfnamefont {S.}~\bibnamefont {Pustelny}}, \bibinfo {author} {\bibfnamefont {A.}~\bibnamefont {Wickenbrock}},\ and\ \bibinfo {author} {\bibfnamefont {A.}~\bibnamefont {Derevianko}},\ }\bibfield  {title} {\bibinfo {title} {Quantum sensor networks as exotic field telescopes for multi-messenger astronomy},\ }\href {https://doi.org/https://doi.org/10.1038/s41550-020-01242-7} {\bibfield  {journal} {\bibinfo  {journal} {Nat. Astron.}\ }\textbf {\bibinfo {volume} {5}},\ \bibinfo {pages} {150} (\bibinfo {year} {2021})}\BibitemShut {NoStop}%
\bibitem [{\citenamefont {Afach}\ \emph {et~al.}(2021)\citenamefont {Afach}, \citenamefont {Buchler}, \citenamefont {Budker}, \citenamefont {Dailey}, \citenamefont {Derevianko}, \citenamefont {Dumont}, \citenamefont {Figueroa}, \citenamefont {Gerhardt}, \citenamefont {Gruji{\'c}}, \citenamefont {Guo} \emph {et~al.}}]{afach2021search}%
  \BibitemOpen
  \bibfield  {author} {\bibinfo {author} {\bibfnamefont {S.}~\bibnamefont {Afach}}, \bibinfo {author} {\bibfnamefont {B.~C.}\ \bibnamefont {Buchler}}, \bibinfo {author} {\bibfnamefont {D.}~\bibnamefont {Budker}}, \bibinfo {author} {\bibfnamefont {C.}~\bibnamefont {Dailey}}, \bibinfo {author} {\bibfnamefont {A.}~\bibnamefont {Derevianko}}, \bibinfo {author} {\bibfnamefont {V.}~\bibnamefont {Dumont}}, \bibinfo {author} {\bibfnamefont {N.~L.}\ \bibnamefont {Figueroa}}, \bibinfo {author} {\bibfnamefont {I.}~\bibnamefont {Gerhardt}}, \bibinfo {author} {\bibfnamefont {Z.~D.}\ \bibnamefont {Gruji{\'c}}}, \bibinfo {author} {\bibfnamefont {H.}~\bibnamefont {Guo}}, \emph {et~al.},\ }\bibfield  {title} {\bibinfo {title} {Search for topological defect dark matter with a global network of optical magnetometers},\ }\href {https://doi.org/https://doi.org/10.1038/s41567-021-01393-y} {\bibfield  {journal} {\bibinfo  {journal} {Nat. Phys.}\ }\textbf {\bibinfo {volume} {17}},\ \bibinfo {pages} {1396} (\bibinfo {year}
  {2021})}\BibitemShut {NoStop}%
\bibitem [{\citenamefont {Jiang}\ \emph {et~al.}(2021)\citenamefont {Jiang}, \citenamefont {Su}, \citenamefont {Garcon}, \citenamefont {Peng},\ and\ \citenamefont {Budker}}]{jiang2021search}%
  \BibitemOpen
  \bibfield  {author} {\bibinfo {author} {\bibfnamefont {M.}~\bibnamefont {Jiang}}, \bibinfo {author} {\bibfnamefont {H.}~\bibnamefont {Su}}, \bibinfo {author} {\bibfnamefont {A.}~\bibnamefont {Garcon}}, \bibinfo {author} {\bibfnamefont {X.}~\bibnamefont {Peng}},\ and\ \bibinfo {author} {\bibfnamefont {D.}~\bibnamefont {Budker}},\ }\bibfield  {title} {\bibinfo {title} {Search for axion-like dark matter with spin-based amplifiers},\ }\href {https://doi.org/https://doi.org/10.1038/s41567-021-01392-z} {\bibfield  {journal} {\bibinfo  {journal} {Nat. Phys.}\ }\textbf {\bibinfo {volume} {17}},\ \bibinfo {pages} {1402} (\bibinfo {year} {2021})}\BibitemShut {NoStop}%
\bibitem [{\citenamefont {Nichol}\ \emph {et~al.}(2022)\citenamefont {Nichol}, \citenamefont {Srinivas}, \citenamefont {Nadlinger}, \citenamefont {Drmota}, \citenamefont {Main}, \citenamefont {Araneda}, \citenamefont {Ballance},\ and\ \citenamefont {Lucas}}]{nichol2022elementary}%
  \BibitemOpen
  \bibfield  {author} {\bibinfo {author} {\bibfnamefont {B.~C.}\ \bibnamefont {Nichol}}, \bibinfo {author} {\bibfnamefont {R.}~\bibnamefont {Srinivas}}, \bibinfo {author} {\bibfnamefont {D.}~\bibnamefont {Nadlinger}}, \bibinfo {author} {\bibfnamefont {P.}~\bibnamefont {Drmota}}, \bibinfo {author} {\bibfnamefont {D.}~\bibnamefont {Main}}, \bibinfo {author} {\bibfnamefont {G.}~\bibnamefont {Araneda}}, \bibinfo {author} {\bibfnamefont {C.}~\bibnamefont {Ballance}},\ and\ \bibinfo {author} {\bibfnamefont {D.}~\bibnamefont {Lucas}},\ }\bibfield  {title} {\bibinfo {title} {An elementary quantum network of entangled optical atomic clocks},\ }\href {https://doi.org/https://doi.org/10.1038/s41586-022-05088-z} {\bibfield  {journal} {\bibinfo  {journal} {Nature}\ }\textbf {\bibinfo {volume} {609}},\ \bibinfo {pages} {689} (\bibinfo {year} {2022})}\BibitemShut {NoStop}%
\bibitem [{\citenamefont {Venema}\ \emph {et~al.}(1992)\citenamefont {Venema}, \citenamefont {Majumder}, \citenamefont {Lamoreaux}, \citenamefont {Heckel},\ and\ \citenamefont {Fortson}}]{venema1992search}%
  \BibitemOpen
  \bibfield  {author} {\bibinfo {author} {\bibfnamefont {B.}~\bibnamefont {Venema}}, \bibinfo {author} {\bibfnamefont {P.}~\bibnamefont {Majumder}}, \bibinfo {author} {\bibfnamefont {S.}~\bibnamefont {Lamoreaux}}, \bibinfo {author} {\bibfnamefont {B.}~\bibnamefont {Heckel}},\ and\ \bibinfo {author} {\bibfnamefont {E.}~\bibnamefont {Fortson}},\ }\bibfield  {title} {\bibinfo {title} {Search for a coupling of the \text{E}arth’s gravitational field to nuclear spins in atomic mercury},\ }\href {https://doi.org/10.1103/PhysRevLett.68.135} {\bibfield  {journal} {\bibinfo  {journal} {Phys. Rev. Lett.}\ }\textbf {\bibinfo {volume} {68}},\ \bibinfo {pages} {135} (\bibinfo {year} {1992})}\BibitemShut {NoStop}%
\bibitem [{\citenamefont {Terrano}\ and\ \citenamefont {Romalis}(2021)}]{terrano2021comagnetometer}%
  \BibitemOpen
  \bibfield  {author} {\bibinfo {author} {\bibfnamefont {W.}~\bibnamefont {Terrano}}\ and\ \bibinfo {author} {\bibfnamefont {M.}~\bibnamefont {Romalis}},\ }\bibfield  {title} {\bibinfo {title} {Comagnetometer probes of dark matter and new physics},\ }\href {https://doi.org/10.1088/2058-9565/ac1ae0} {\bibfield  {journal} {\bibinfo  {journal} {Quantum Sci. Technol.}\ }\textbf {\bibinfo {volume} {7}},\ \bibinfo {pages} {014001} (\bibinfo {year} {2021})}\BibitemShut {NoStop}%
\bibitem [{\citenamefont {Jackson~Kimball}\ \emph {et~al.}(2017)\citenamefont {Jackson~Kimball}, \citenamefont {Dudley}, \citenamefont {Li}, \citenamefont {Patel},\ and\ \citenamefont {Valdez}}]{jackson2017constraints}%
  \BibitemOpen
  \bibfield  {author} {\bibinfo {author} {\bibfnamefont {D.~F.}\ \bibnamefont {Jackson~Kimball}}, \bibinfo {author} {\bibfnamefont {J.}~\bibnamefont {Dudley}}, \bibinfo {author} {\bibfnamefont {Y.}~\bibnamefont {Li}}, \bibinfo {author} {\bibfnamefont {D.}~\bibnamefont {Patel}},\ and\ \bibinfo {author} {\bibfnamefont {J.}~\bibnamefont {Valdez}},\ }\bibfield  {title} {\bibinfo {title} {Constraints on long-range spin-gravity and monopole-dipole couplings of the proton},\ }\href {https://doi.org/10.1103/PhysRevD.96.075004} {\bibfield  {journal} {\bibinfo  {journal} {Phys. Rev. D}\ }\textbf {\bibinfo {volume} {96}},\ \bibinfo {pages} {075004} (\bibinfo {year} {2017})}\BibitemShut {NoStop}%
\bibitem [{\citenamefont {Zhang}\ \emph {et~al.}(2023{\natexlab{a}})\citenamefont {Zhang}, \citenamefont {Ba}, \citenamefont {Ning}, \citenamefont {Zhai}, \citenamefont {Lu},\ and\ \citenamefont {Sheng}}]{ShengDong2023SG}%
  \BibitemOpen
  \bibfield  {author} {\bibinfo {author} {\bibfnamefont {S.-B.}\ \bibnamefont {Zhang}}, \bibinfo {author} {\bibfnamefont {Z.-L.}\ \bibnamefont {Ba}}, \bibinfo {author} {\bibfnamefont {D.-H.}\ \bibnamefont {Ning}}, \bibinfo {author} {\bibfnamefont {N.-F.}\ \bibnamefont {Zhai}}, \bibinfo {author} {\bibfnamefont {Z.-T.}\ \bibnamefont {Lu}},\ and\ \bibinfo {author} {\bibfnamefont {D.}~\bibnamefont {Sheng}},\ }\bibfield  {title} {\bibinfo {title} {Search for spin-dependent gravitational interactions at earth range},\ }\href {https://doi.org/10.1103/PhysRevLett.130.201401} {\bibfield  {journal} {\bibinfo  {journal} {Phys. Rev. Lett.}\ }\textbf {\bibinfo {volume} {130}},\ \bibinfo {pages} {201401} (\bibinfo {year} {2023}{\natexlab{a}})}\BibitemShut {NoStop}%
\bibitem [{\citenamefont {Rosenberry}\ and\ \citenamefont {Chupp}(2001)}]{rosenberry2001atomic}%
  \BibitemOpen
  \bibfield  {author} {\bibinfo {author} {\bibfnamefont {M.}~\bibnamefont {Rosenberry}}\ and\ \bibinfo {author} {\bibfnamefont {T.}~\bibnamefont {Chupp}},\ }\bibfield  {title} {\bibinfo {title} {Atomic electric dipole moment measurement using spin exchange pumped masers of $^{129}\text{Xe}$ and $^3\text{He}$},\ }\href {https://doi.org/10.1103/PhysRevLett.86.22} {\bibfield  {journal} {\bibinfo  {journal} {Phys. Rev. Lett.}\ }\textbf {\bibinfo {volume} {86}},\ \bibinfo {pages} {22} (\bibinfo {year} {2001})}\BibitemShut {NoStop}%
\bibitem [{\citenamefont {Baker}\ \emph {et~al.}(2006)\citenamefont {Baker}, \citenamefont {Doyle}, \citenamefont {Geltenbort}, \citenamefont {Green}, \citenamefont {Van~der Grinten}, \citenamefont {Harris}, \citenamefont {Iaydjiev}, \citenamefont {Ivanov}, \citenamefont {May}, \citenamefont {Pendlebury} \emph {et~al.}}]{baker2006improved}%
  \BibitemOpen
  \bibfield  {author} {\bibinfo {author} {\bibfnamefont {C.}~\bibnamefont {Baker}}, \bibinfo {author} {\bibfnamefont {D.}~\bibnamefont {Doyle}}, \bibinfo {author} {\bibfnamefont {P.}~\bibnamefont {Geltenbort}}, \bibinfo {author} {\bibfnamefont {K.}~\bibnamefont {Green}}, \bibinfo {author} {\bibfnamefont {M.}~\bibnamefont {Van~der Grinten}}, \bibinfo {author} {\bibfnamefont {P.}~\bibnamefont {Harris}}, \bibinfo {author} {\bibfnamefont {P.}~\bibnamefont {Iaydjiev}}, \bibinfo {author} {\bibfnamefont {S.}~\bibnamefont {Ivanov}}, \bibinfo {author} {\bibfnamefont {D.}~\bibnamefont {May}}, \bibinfo {author} {\bibfnamefont {J.}~\bibnamefont {Pendlebury}}, \emph {et~al.},\ }\bibfield  {title} {\bibinfo {title} {Improved experimental limit on the electric dipole moment of the neutron},\ }\href {https://doi.org/10.1103/PhysRevLett.97.131801} {\bibfield  {journal} {\bibinfo  {journal} {Phys. Rev. Lett.}\ }\textbf {\bibinfo {volume} {97}},\ \bibinfo {pages} {131801} (\bibinfo {year} {2006})}\BibitemShut {NoStop}%
\bibitem [{\citenamefont {Roberts}\ \emph {et~al.}(2017)\citenamefont {Roberts}, \citenamefont {Blewitt}, \citenamefont {Dailey}, \citenamefont {Murphy}, \citenamefont {Pospelov}, \citenamefont {Rollings}, \citenamefont {Sherman}, \citenamefont {Williams},\ and\ \citenamefont {Derevianko}}]{roberts2017search}%
  \BibitemOpen
  \bibfield  {author} {\bibinfo {author} {\bibfnamefont {B.~M.}\ \bibnamefont {Roberts}}, \bibinfo {author} {\bibfnamefont {G.}~\bibnamefont {Blewitt}}, \bibinfo {author} {\bibfnamefont {C.}~\bibnamefont {Dailey}}, \bibinfo {author} {\bibfnamefont {M.}~\bibnamefont {Murphy}}, \bibinfo {author} {\bibfnamefont {M.}~\bibnamefont {Pospelov}}, \bibinfo {author} {\bibfnamefont {A.}~\bibnamefont {Rollings}}, \bibinfo {author} {\bibfnamefont {J.}~\bibnamefont {Sherman}}, \bibinfo {author} {\bibfnamefont {W.}~\bibnamefont {Williams}},\ and\ \bibinfo {author} {\bibfnamefont {A.}~\bibnamefont {Derevianko}},\ }\bibfield  {title} {\bibinfo {title} {Search for domain wall dark matter with atomic clocks on board global positioning system satellites},\ }\href {https://doi.org/https://doi.org/10.1038/s41467-017-01440-4} {\bibfield  {journal} {\bibinfo  {journal} {Nat. Commun.}\ }\textbf {\bibinfo {volume} {8}},\ \bibinfo {pages} {1195} (\bibinfo {year} {2017})}\BibitemShut {NoStop}%
\bibitem [{\citenamefont {Jackson~Kimball}\ \emph {et~al.}(2018)\citenamefont {Jackson~Kimball}, \citenamefont {Budker}, \citenamefont {Eby}, \citenamefont {Pospelov}, \citenamefont {Pustelny}, \citenamefont {Scholtes}, \citenamefont {Stadnik}, \citenamefont {Weis},\ and\ \citenamefont {Wickenbrock}}]{jackson2018searching}%
  \BibitemOpen
  \bibfield  {author} {\bibinfo {author} {\bibfnamefont {D.~F.}\ \bibnamefont {Jackson~Kimball}}, \bibinfo {author} {\bibfnamefont {D.}~\bibnamefont {Budker}}, \bibinfo {author} {\bibfnamefont {J.}~\bibnamefont {Eby}}, \bibinfo {author} {\bibfnamefont {M.}~\bibnamefont {Pospelov}}, \bibinfo {author} {\bibfnamefont {S.}~\bibnamefont {Pustelny}}, \bibinfo {author} {\bibfnamefont {T.}~\bibnamefont {Scholtes}}, \bibinfo {author} {\bibfnamefont {Y.}~\bibnamefont {Stadnik}}, \bibinfo {author} {\bibfnamefont {A.}~\bibnamefont {Weis}},\ and\ \bibinfo {author} {\bibfnamefont {A.}~\bibnamefont {Wickenbrock}},\ }\bibfield  {title} {\bibinfo {title} {Searching for axion stars and $q$-balls with a terrestrial magnetometer network},\ }\href {https://doi.org/10.1103/PhysRevD.97.043002} {\bibfield  {journal} {\bibinfo  {journal} {Phys. Rev. D}\ }\textbf {\bibinfo {volume} {97}},\ \bibinfo {pages} {043002} (\bibinfo {year} {2018})}\BibitemShut {NoStop}%
\bibitem [{\citenamefont {Graham}\ and\ \citenamefont {Rajendran}(2013)}]{graham2013new}%
  \BibitemOpen
  \bibfield  {author} {\bibinfo {author} {\bibfnamefont {P.~W.}\ \bibnamefont {Graham}}\ and\ \bibinfo {author} {\bibfnamefont {S.}~\bibnamefont {Rajendran}},\ }\bibfield  {title} {\bibinfo {title} {New observables for direct detection of axion dark matter},\ }\href {https://doi.org/10.1103/PhysRevD.88.035023} {\bibfield  {journal} {\bibinfo  {journal} {Phys. Rev. D}\ }\textbf {\bibinfo {volume} {88}},\ \bibinfo {pages} {035023} (\bibinfo {year} {2013})}\BibitemShut {NoStop}%
\bibitem [{\citenamefont {Gao}\ \emph {et~al.}(2022)\citenamefont {Gao}, \citenamefont {Halperin}, \citenamefont {Kahn}, \citenamefont {Nguyen}, \citenamefont {Sch{\"u}tte-Engel},\ and\ \citenamefont {Scott}}]{gao2022axion}%
  \BibitemOpen
  \bibfield  {author} {\bibinfo {author} {\bibfnamefont {C.}~\bibnamefont {Gao}}, \bibinfo {author} {\bibfnamefont {W.}~\bibnamefont {Halperin}}, \bibinfo {author} {\bibfnamefont {Y.}~\bibnamefont {Kahn}}, \bibinfo {author} {\bibfnamefont {M.}~\bibnamefont {Nguyen}}, \bibinfo {author} {\bibfnamefont {J.}~\bibnamefont {Sch{\"u}tte-Engel}},\ and\ \bibinfo {author} {\bibfnamefont {J.~W.}\ \bibnamefont {Scott}},\ }\bibfield  {title} {\bibinfo {title} {Axion wind detection with the homogeneous precession domain of superfluid helium-3},\ }\href {https://doi.org/10.1103/PhysRevLett.129.211801} {\bibfield  {journal} {\bibinfo  {journal} {Phys. Rev. Lett.}\ }\textbf {\bibinfo {volume} {129}},\ \bibinfo {pages} {211801} (\bibinfo {year} {2022})}\BibitemShut {NoStop}%
\bibitem [{\citenamefont {Tullney}\ \emph {et~al.}(2013)\citenamefont {Tullney}, \citenamefont {Allmendinger}, \citenamefont {Burghoff}, \citenamefont {Heil}, \citenamefont {Karpuk}, \citenamefont {Kilian}, \citenamefont {Knappe-Gr{\"u}neberg}, \citenamefont {M{\"u}ller}, \citenamefont {Schmidt}, \citenamefont {Schnabel} \emph {et~al.}}]{tullney2013constraints}%
  \BibitemOpen
  \bibfield  {author} {\bibinfo {author} {\bibfnamefont {K.}~\bibnamefont {Tullney}}, \bibinfo {author} {\bibfnamefont {F.}~\bibnamefont {Allmendinger}}, \bibinfo {author} {\bibfnamefont {M.}~\bibnamefont {Burghoff}}, \bibinfo {author} {\bibfnamefont {W.}~\bibnamefont {Heil}}, \bibinfo {author} {\bibfnamefont {S.}~\bibnamefont {Karpuk}}, \bibinfo {author} {\bibfnamefont {W.}~\bibnamefont {Kilian}}, \bibinfo {author} {\bibfnamefont {S.}~\bibnamefont {Knappe-Gr{\"u}neberg}}, \bibinfo {author} {\bibfnamefont {W.}~\bibnamefont {M{\"u}ller}}, \bibinfo {author} {\bibfnamefont {U.}~\bibnamefont {Schmidt}}, \bibinfo {author} {\bibfnamefont {A.}~\bibnamefont {Schnabel}}, \emph {et~al.},\ }\bibfield  {title} {\bibinfo {title} {Constraints on spin-dependent short-range interaction between nucleons},\ }\href {https://doi.org/10.1103/PhysRevLett.111.100801} {\bibfield  {journal} {\bibinfo  {journal} {Phys. Rev. Lett.}\ }\textbf {\bibinfo {volume} {111}},\ \bibinfo {pages} {100801} (\bibinfo {year} {2013})}\BibitemShut
  {NoStop}%
\bibitem [{\citenamefont {Wang}\ \emph {et~al.}(2020)\citenamefont {Wang}, \citenamefont {Peng}, \citenamefont {Zhang}, \citenamefont {Luo}, \citenamefont {Li}, \citenamefont {Xiong}, \citenamefont {Wang},\ and\ \citenamefont {Guo}}]{wang2020single}%
  \BibitemOpen
  \bibfield  {author} {\bibinfo {author} {\bibfnamefont {Z.}~\bibnamefont {Wang}}, \bibinfo {author} {\bibfnamefont {X.}~\bibnamefont {Peng}}, \bibinfo {author} {\bibfnamefont {R.}~\bibnamefont {Zhang}}, \bibinfo {author} {\bibfnamefont {H.}~\bibnamefont {Luo}}, \bibinfo {author} {\bibfnamefont {J.}~\bibnamefont {Li}}, \bibinfo {author} {\bibfnamefont {Z.}~\bibnamefont {Xiong}}, \bibinfo {author} {\bibfnamefont {S.}~\bibnamefont {Wang}},\ and\ \bibinfo {author} {\bibfnamefont {H.}~\bibnamefont {Guo}},\ }\bibfield  {title} {\bibinfo {title} {Single-species atomic comagnetometer based on $^{87}\text{Rb}$ atoms},\ }\href {https://doi.org/10.1103/PhysRevLett.124.193002} {\bibfield  {journal} {\bibinfo  {journal} {Phys. Rev. Lett.}\ }\textbf {\bibinfo {volume} {124}},\ \bibinfo {pages} {193002} (\bibinfo {year} {2020})}\BibitemShut {NoStop}%
\bibitem [{\citenamefont {Zhang}\ \emph {et~al.}(2023{\natexlab{b}})\citenamefont {Zhang}, \citenamefont {Ba}, \citenamefont {Ning}, \citenamefont {Zhai}, \citenamefont {Lu},\ and\ \citenamefont {Sheng}}]{zhang2023search}%
  \BibitemOpen
  \bibfield  {author} {\bibinfo {author} {\bibfnamefont {S.-B.}\ \bibnamefont {Zhang}}, \bibinfo {author} {\bibfnamefont {Z.-L.}\ \bibnamefont {Ba}}, \bibinfo {author} {\bibfnamefont {D.-H.}\ \bibnamefont {Ning}}, \bibinfo {author} {\bibfnamefont {N.-F.}\ \bibnamefont {Zhai}}, \bibinfo {author} {\bibfnamefont {Z.-T.}\ \bibnamefont {Lu}},\ and\ \bibinfo {author} {\bibfnamefont {D.}~\bibnamefont {Sheng}},\ }\bibfield  {title} {\bibinfo {title} {Search for spin-dependent gravitational interactions at \text{Earth} range},\ }\href {https://doi.org/10.1103/PhysRevLett.130.201401} {\bibfield  {journal} {\bibinfo  {journal} {Phys. Rev. Lett.}\ }\textbf {\bibinfo {volume} {130}},\ \bibinfo {pages} {201401} (\bibinfo {year} {2023}{\natexlab{b}})}\BibitemShut {NoStop}%
\bibitem [{\citenamefont {Kornack}\ and\ \citenamefont {Romalis}(2002)}]{kornack2002dynamics}%
  \BibitemOpen
  \bibfield  {author} {\bibinfo {author} {\bibfnamefont {T.~W.}\ \bibnamefont {Kornack}}\ and\ \bibinfo {author} {\bibfnamefont {M.~V.}\ \bibnamefont {Romalis}},\ }\bibfield  {title} {\bibinfo {title} {Dynamics of two overlapping spin ensembles interacting by spin exchange},\ }\href {https://doi.org/10.1103/PhysRevLett.89.253002} {\bibfield  {journal} {\bibinfo  {journal} {Phys. Rev. Lett.}\ }\textbf {\bibinfo {volume} {89}},\ \bibinfo {pages} {253002} (\bibinfo {year} {2002})}\BibitemShut {NoStop}%
\bibitem [{\citenamefont {Lee}\ \emph {et~al.}(2018)\citenamefont {Lee}, \citenamefont {Almasi},\ and\ \citenamefont {Romalis}}]{lee2018improved}%
  \BibitemOpen
  \bibfield  {author} {\bibinfo {author} {\bibfnamefont {J.}~\bibnamefont {Lee}}, \bibinfo {author} {\bibfnamefont {A.}~\bibnamefont {Almasi}},\ and\ \bibinfo {author} {\bibfnamefont {M.}~\bibnamefont {Romalis}},\ }\bibfield  {title} {\bibinfo {title} {Improved limits on spin-mass interactions},\ }\href {https://doi.org/10.1103/PhysRevLett.120.161801} {\bibfield  {journal} {\bibinfo  {journal} {Phys. Rev. Lett.}\ }\textbf {\bibinfo {volume} {120}},\ \bibinfo {pages} {161801} (\bibinfo {year} {2018})}\BibitemShut {NoStop}%
\bibitem [{\citenamefont {Padniuk}\ \emph {et~al.}(2022)\citenamefont {Padniuk}, \citenamefont {Kopciuch}, \citenamefont {Cipolletti}, \citenamefont {Wickenbrock}, \citenamefont {Budker},\ and\ \citenamefont {Pustelny}}]{padniuk2022response}%
  \BibitemOpen
  \bibfield  {author} {\bibinfo {author} {\bibfnamefont {M.}~\bibnamefont {Padniuk}}, \bibinfo {author} {\bibfnamefont {M.}~\bibnamefont {Kopciuch}}, \bibinfo {author} {\bibfnamefont {R.}~\bibnamefont {Cipolletti}}, \bibinfo {author} {\bibfnamefont {A.}~\bibnamefont {Wickenbrock}}, \bibinfo {author} {\bibfnamefont {D.}~\bibnamefont {Budker}},\ and\ \bibinfo {author} {\bibfnamefont {S.}~\bibnamefont {Pustelny}},\ }\bibfield  {title} {\bibinfo {title} {Response of atomic spin-based sensors to magnetic and nonmagnetic perturbations},\ }\href {https://doi.org/https://doi.org/10.1038/s41598-021-03609-w} {\bibfield  {journal} {\bibinfo  {journal} {Sci. Rep.}\ }\textbf {\bibinfo {volume} {12}},\ \bibinfo {pages} {324} (\bibinfo {year} {2022})}\BibitemShut {NoStop}%
\bibitem [{\citenamefont {Padniuk}\ \emph {et~al.}(2024)\citenamefont {Padniuk}, \citenamefont {Klinger}, \citenamefont {{\L}ukasiewicz}, \citenamefont {Gavilan-Martin}, \citenamefont {Liu}, \citenamefont {Pustelny}, \citenamefont {Jackson~Kimball}, \citenamefont {Budker},\ and\ \citenamefont {Wickenbrock}}]{padniuk2024universal}%
  \BibitemOpen
  \bibfield  {author} {\bibinfo {author} {\bibfnamefont {M.}~\bibnamefont {Padniuk}}, \bibinfo {author} {\bibfnamefont {E.}~\bibnamefont {Klinger}}, \bibinfo {author} {\bibfnamefont {G.}~\bibnamefont {{\L}ukasiewicz}}, \bibinfo {author} {\bibfnamefont {D.}~\bibnamefont {Gavilan-Martin}}, \bibinfo {author} {\bibfnamefont {T.}~\bibnamefont {Liu}}, \bibinfo {author} {\bibfnamefont {S.}~\bibnamefont {Pustelny}}, \bibinfo {author} {\bibfnamefont {D.~F.}\ \bibnamefont {Jackson~Kimball}}, \bibinfo {author} {\bibfnamefont {D.}~\bibnamefont {Budker}},\ and\ \bibinfo {author} {\bibfnamefont {A.}~\bibnamefont {Wickenbrock}},\ }\bibfield  {title} {\bibinfo {title} {Universal determination of comagnetometer response to spin couplings},\ }\href {https://doi.org/10.1103/PhysRevResearch.6.013339} {\bibfield  {journal} {\bibinfo  {journal} {Phys. Rev. Research}\ }\textbf {\bibinfo {volume} {6}},\ \bibinfo {pages} {013339} (\bibinfo {year} {2024})}\BibitemShut {NoStop}%
\bibitem [{\citenamefont {Rosenzweig}\ \emph {et~al.}(2024)\citenamefont {Rosenzweig}, \citenamefont {Kats}, \citenamefont {Givon}, \citenamefont {Japha},\ and\ \citenamefont {Folman}}]{rosenzweig2024atomic}%
  \BibitemOpen
  \bibfield  {author} {\bibinfo {author} {\bibfnamefont {Y.}~\bibnamefont {Rosenzweig}}, \bibinfo {author} {\bibfnamefont {Y.}~\bibnamefont {Kats}}, \bibinfo {author} {\bibfnamefont {M.}~\bibnamefont {Givon}}, \bibinfo {author} {\bibfnamefont {Y.}~\bibnamefont {Japha}},\ and\ \bibinfo {author} {\bibfnamefont {R.}~\bibnamefont {Folman}},\ }\bibfield  {title} {\bibinfo {title} {Atomic probe of dark matter differential interactions with subatomic particles},\ }\href {https://doi.org/10.1103/PhysRevD.110.015015} {\bibfield  {journal} {\bibinfo  {journal} {Phys. Rev. D}\ }\textbf {\bibinfo {volume} {110}},\ \bibinfo {pages} {015015} (\bibinfo {year} {2024})}\BibitemShut {NoStop}%
\bibitem [{\citenamefont {Zhao}\ \emph {et~al.}(2024)\citenamefont {Zhao}, \citenamefont {Zheng}, \citenamefont {Xiao}, \citenamefont {Peng}, \citenamefont {Wu},\ and\ \citenamefont {Guo}}]{zhao2024suppression}%
  \BibitemOpen
  \bibfield  {author} {\bibinfo {author} {\bibfnamefont {Y.}~\bibnamefont {Zhao}}, \bibinfo {author} {\bibfnamefont {J.}~\bibnamefont {Zheng}}, \bibinfo {author} {\bibfnamefont {W.}~\bibnamefont {Xiao}}, \bibinfo {author} {\bibfnamefont {X.}~\bibnamefont {Peng}}, \bibinfo {author} {\bibfnamefont {T.}~\bibnamefont {Wu}},\ and\ \bibinfo {author} {\bibfnamefont {H.}~\bibnamefont {Guo}},\ }\bibfield  {title} {\bibinfo {title} {Suppression of systematic effects in a single-species atomic comagnetometer},\ }\href {https://doi.org/10.1103/PhysRevA.110.063122} {\bibfield  {journal} {\bibinfo  {journal} {Phys. Rev. A}\ }\textbf {\bibinfo {volume} {110}},\ \bibinfo {pages} {063122} (\bibinfo {year} {2024})}\BibitemShut {NoStop}%
\bibitem [{\citenamefont {Sheng}\ \emph {et~al.}(2017)\citenamefont {Sheng}, \citenamefont {Perry}, \citenamefont {Krzyzewski}, \citenamefont {Geller}, \citenamefont {Kitching},\ and\ \citenamefont {Knappe}}]{sheng2017microfabricated}%
  \BibitemOpen
  \bibfield  {author} {\bibinfo {author} {\bibfnamefont {D.}~\bibnamefont {Sheng}}, \bibinfo {author} {\bibfnamefont {A.~R.}\ \bibnamefont {Perry}}, \bibinfo {author} {\bibfnamefont {S.~P.}\ \bibnamefont {Krzyzewski}}, \bibinfo {author} {\bibfnamefont {S.}~\bibnamefont {Geller}}, \bibinfo {author} {\bibfnamefont {J.}~\bibnamefont {Kitching}},\ and\ \bibinfo {author} {\bibfnamefont {S.}~\bibnamefont {Knappe}},\ }\bibfield  {title} {\bibinfo {title} {A microfabricated optically-pumped magnetic gradiometer},\ }\href {https://doi.org/https://doi.org/10.1063/1.4974349} {\bibfield  {journal} {\bibinfo  {journal} {Appl. Phys. Lett.}\ }\textbf {\bibinfo {volume} {110}},\ \bibinfo {pages} {031106} (\bibinfo {year} {2017})}\BibitemShut {NoStop}%
\bibitem [{\citenamefont {Sulai}\ \emph {et~al.}(2019)\citenamefont {Sulai}, \citenamefont {DeLand}, \citenamefont {Bulatowicz}, \citenamefont {Wahl}, \citenamefont {Wakai},\ and\ \citenamefont {Walker}}]{sulai2019characterizing}%
  \BibitemOpen
  \bibfield  {author} {\bibinfo {author} {\bibfnamefont {I.}~\bibnamefont {Sulai}}, \bibinfo {author} {\bibfnamefont {Z.}~\bibnamefont {DeLand}}, \bibinfo {author} {\bibfnamefont {M.}~\bibnamefont {Bulatowicz}}, \bibinfo {author} {\bibfnamefont {C.}~\bibnamefont {Wahl}}, \bibinfo {author} {\bibfnamefont {R.}~\bibnamefont {Wakai}},\ and\ \bibinfo {author} {\bibfnamefont {T.}~\bibnamefont {Walker}},\ }\bibfield  {title} {\bibinfo {title} {Characterizing atomic magnetic gradiometers for fetal magnetocardiography},\ }\href {https://doi.org/https://doi.org/10.1063/1.5091007} {\bibfield  {journal} {\bibinfo  {journal} {Rev. Sci. Instrum.}\ }\textbf {\bibinfo {volume} {90}},\ \bibinfo {pages} {085003} (\bibinfo {year} {2019})}\BibitemShut {NoStop}%
\bibitem [{\citenamefont {Dong}\ \emph {et~al.}(2023)\citenamefont {Dong}, \citenamefont {Ye}, \citenamefont {Hu},\ and\ \citenamefont {Ma}}]{dong2023recent}%
  \BibitemOpen
  \bibfield  {author} {\bibinfo {author} {\bibfnamefont {H.}~\bibnamefont {Dong}}, \bibinfo {author} {\bibfnamefont {H.}~\bibnamefont {Ye}}, \bibinfo {author} {\bibfnamefont {M.}~\bibnamefont {Hu}},\ and\ \bibinfo {author} {\bibfnamefont {Z.}~\bibnamefont {Ma}},\ }\bibfield  {title} {\bibinfo {title} {Recent developments in fabrication methods and measurement schemes for optically pumped magnetic gradiometers: a comprehensive review},\ }\href {https://doi.org/https://doi.org/10.3390/mi15010059} {\bibfield  {journal} {\bibinfo  {journal} {Micromachines}\ }\textbf {\bibinfo {volume} {15}},\ \bibinfo {pages} {59} (\bibinfo {year} {2023})}\BibitemShut {NoStop}%
\bibitem [{\citenamefont {Peccei}\ and\ \citenamefont {Quinn}(1977)}]{peccei1977constraints}%
  \BibitemOpen
  \bibfield  {author} {\bibinfo {author} {\bibfnamefont {R.~D.}\ \bibnamefont {Peccei}}\ and\ \bibinfo {author} {\bibfnamefont {H.~R.}\ \bibnamefont {Quinn}},\ }\bibfield  {title} {\bibinfo {title} {Constraints imposed by $\mathrm{CP}$ conservation in the presence of pseudoparticles},\ }\href {https://doi.org/10.1103/PhysRevD.16.1791} {\bibfield  {journal} {\bibinfo  {journal} {Phys. Rev. D}\ }\textbf {\bibinfo {volume} {16}},\ \bibinfo {pages} {1791} (\bibinfo {year} {1977})}\BibitemShut {NoStop}%
\bibitem [{\citenamefont {Schmidt}(1937)}]{schmidt1937magnetischen}%
  \BibitemOpen
  \bibfield  {author} {\bibinfo {author} {\bibfnamefont {T.}~\bibnamefont {Schmidt}},\ }\bibfield  {title} {\bibinfo {title} {Über die magnetischen momente der atomkerne},\ }\href {https://doi.org/https://doi.org/10.1007/BF01339933} {\bibfield  {journal} {\bibinfo  {journal} {Z. Physik}\ }\textbf {\bibinfo {volume} {106}},\ \bibinfo {pages} {358} (\bibinfo {year} {1937})}\BibitemShut {NoStop}%
\bibitem [{\citenamefont {Jackson~Kimball}(2015)}]{kimball2015nuclear}%
  \BibitemOpen
  \bibfield  {author} {\bibinfo {author} {\bibfnamefont {D.~F.}\ \bibnamefont {Jackson~Kimball}},\ }\bibfield  {title} {\bibinfo {title} {Nuclear spin content and constraints on exotic spin-dependent couplings},\ }\href {https://doi.org/10.1088/1367-2630/17/7/073008} {\bibfield  {journal} {\bibinfo  {journal} {New J. Phys.}\ }\textbf {\bibinfo {volume} {17}},\ \bibinfo {pages} {073008} (\bibinfo {year} {2015})}\BibitemShut {NoStop}%
\bibitem [{\citenamefont {Stadnik}\ and\ \citenamefont {Flambaum}(2014)}]{stadnik2014axion}%
  \BibitemOpen
  \bibfield  {author} {\bibinfo {author} {\bibfnamefont {Y.~V.}\ \bibnamefont {Stadnik}}\ and\ \bibinfo {author} {\bibfnamefont {V.}~\bibnamefont {Flambaum}},\ }\bibfield  {title} {\bibinfo {title} {Axion-induced effects in atoms, molecules, and nuclei: Parity nonconservation, anapole moments, electric dipole moments, and spin-gravity and spin-axion momentum couplings},\ }\href {https://doi.org/https://doi.org/10.1103/PhysRevD.89.043522} {\bibfield  {journal} {\bibinfo  {journal} {Phys. Rev. D}\ }\textbf {\bibinfo {volume} {89}},\ \bibinfo {pages} {043522} (\bibinfo {year} {2014})}\BibitemShut {NoStop}%
\bibitem [{\citenamefont {Wu}\ \emph {et~al.}(2019)\citenamefont {Wu}, \citenamefont {Blanchard}, \citenamefont {Centers}, \citenamefont {Figueroa}, \citenamefont {Garcon}, \citenamefont {Graham}, \citenamefont {Kimball}, \citenamefont {Rajendran}, \citenamefont {Stadnik}, \citenamefont {Sushkov} \emph {et~al.}}]{wu2019search}%
  \BibitemOpen
  \bibfield  {author} {\bibinfo {author} {\bibfnamefont {T.}~\bibnamefont {Wu}}, \bibinfo {author} {\bibfnamefont {J.~W.}\ \bibnamefont {Blanchard}}, \bibinfo {author} {\bibfnamefont {G.~P.}\ \bibnamefont {Centers}}, \bibinfo {author} {\bibfnamefont {N.~L.}\ \bibnamefont {Figueroa}}, \bibinfo {author} {\bibfnamefont {A.}~\bibnamefont {Garcon}}, \bibinfo {author} {\bibfnamefont {P.~W.}\ \bibnamefont {Graham}}, \bibinfo {author} {\bibfnamefont {D.~F.~J.}\ \bibnamefont {Kimball}}, \bibinfo {author} {\bibfnamefont {S.}~\bibnamefont {Rajendran}}, \bibinfo {author} {\bibfnamefont {Y.~V.}\ \bibnamefont {Stadnik}}, \bibinfo {author} {\bibfnamefont {A.~O.}\ \bibnamefont {Sushkov}}, \emph {et~al.},\ }\bibfield  {title} {\bibinfo {title} {Search for axionlike dark matter with a liquid-state nuclear spin comagnetometer},\ }\href {https://doi.org/https://doi.org/10.1103/PhysRevLett.122.191302} {\bibfield  {journal} {\bibinfo  {journal} {Phys. Rev. Lett.}\ }\textbf {\bibinfo {volume} {122}},\ \bibinfo {pages} {191302}
  (\bibinfo {year} {2019})}\BibitemShut {NoStop}%
\bibitem [{\citenamefont {Bevington}\ \emph {et~al.}(2020)\citenamefont {Bevington}, \citenamefont {Gartman}, \citenamefont {Stadnik},\ and\ \citenamefont {Chalupczak}}]{bevington2020dual}%
  \BibitemOpen
  \bibfield  {author} {\bibinfo {author} {\bibfnamefont {P.}~\bibnamefont {Bevington}}, \bibinfo {author} {\bibfnamefont {R.}~\bibnamefont {Gartman}}, \bibinfo {author} {\bibfnamefont {Y.~V.}\ \bibnamefont {Stadnik}},\ and\ \bibinfo {author} {\bibfnamefont {W.}~\bibnamefont {Chalupczak}},\ }\bibfield  {title} {\bibinfo {title} {Dual-frequency cesium spin maser},\ }\href {https://doi.org/https://doi.org/10.1103/PhysRevA.102.032804} {\bibfield  {journal} {\bibinfo  {journal} {Phys. Rev. A}\ }\textbf {\bibinfo {volume} {102}},\ \bibinfo {pages} {032804} (\bibinfo {year} {2020})}\BibitemShut {NoStop}%
\bibitem [{\citenamefont {Wang}\ \emph {et~al.}(2023)\citenamefont {Wang}, \citenamefont {Peng}, \citenamefont {Xiong}, \citenamefont {Zheng}, \citenamefont {Luo},\ and\ \citenamefont {Guo}}]{wang2023atomic}%
  \BibitemOpen
  \bibfield  {author} {\bibinfo {author} {\bibfnamefont {Z.}~\bibnamefont {Wang}}, \bibinfo {author} {\bibfnamefont {X.}~\bibnamefont {Peng}}, \bibinfo {author} {\bibfnamefont {Z.}~\bibnamefont {Xiong}}, \bibinfo {author} {\bibfnamefont {J.}~\bibnamefont {Zheng}}, \bibinfo {author} {\bibfnamefont {H.}~\bibnamefont {Luo}},\ and\ \bibinfo {author} {\bibfnamefont {H.}~\bibnamefont {Guo}},\ }\bibfield  {title} {\bibinfo {title} {Atomic comagnetometer with a closed-loop optically aligned $^{87}\text{Rb}$ dual-frequency oscillator},\ }\href {https://doi.org/10.1103/PhysRevApplied.19.024021} {\bibfield  {journal} {\bibinfo  {journal} {Phys. Rev. Appl.}\ }\textbf {\bibinfo {volume} {19}},\ \bibinfo {pages} {024021} (\bibinfo {year} {2023})}\BibitemShut {NoStop}%
\bibitem [{\citenamefont {Yang}\ \emph {et~al.}(2021)\citenamefont {Yang}, \citenamefont {Wu}, \citenamefont {Chen}, \citenamefont {Peng},\ and\ \citenamefont {Guo}}]{yang2021all}%
  \BibitemOpen
  \bibfield  {author} {\bibinfo {author} {\bibfnamefont {Y.}~\bibnamefont {Yang}}, \bibinfo {author} {\bibfnamefont {T.}~\bibnamefont {Wu}}, \bibinfo {author} {\bibfnamefont {J.}~\bibnamefont {Chen}}, \bibinfo {author} {\bibfnamefont {X.}~\bibnamefont {Peng}},\ and\ \bibinfo {author} {\bibfnamefont {H.}~\bibnamefont {Guo}},\ }\bibfield  {title} {\bibinfo {title} {All-optical single-species cesium atomic comagnetometer with optical free induction decay detection},\ }\href {https://doi.org/https://doi.org/10.1007/s00340-021-07594-w} {\bibfield  {journal} {\bibinfo  {journal} {Appl. Phys. B}\ }\textbf {\bibinfo {volume} {127}},\ \bibinfo {pages} {1} (\bibinfo {year} {2021})}\BibitemShut {NoStop}%
\bibitem [{\citenamefont {Hunter}\ \emph {et~al.}(2018)\citenamefont {Hunter}, \citenamefont {Jim{\'e}nez-Mart{\'\i}nez}, \citenamefont {Herbsommer}, \citenamefont {Ramaswamy}, \citenamefont {Li},\ and\ \citenamefont {Riis}}]{hunter2018waveform}%
  \BibitemOpen
  \bibfield  {author} {\bibinfo {author} {\bibfnamefont {D.}~\bibnamefont {Hunter}}, \bibinfo {author} {\bibfnamefont {R.}~\bibnamefont {Jim{\'e}nez-Mart{\'\i}nez}}, \bibinfo {author} {\bibfnamefont {J.}~\bibnamefont {Herbsommer}}, \bibinfo {author} {\bibfnamefont {S.}~\bibnamefont {Ramaswamy}}, \bibinfo {author} {\bibfnamefont {W.}~\bibnamefont {Li}},\ and\ \bibinfo {author} {\bibfnamefont {E.}~\bibnamefont {Riis}},\ }\bibfield  {title} {\bibinfo {title} {Waveform reconstruction with a \text{Cs} based free-induction-decay magnetometer},\ }\href {https://doi.org/https://doi.org/10.1364/OE.26.030523} {\bibfield  {journal} {\bibinfo  {journal} {Opt. Express}\ }\textbf {\bibinfo {volume} {26}},\ \bibinfo {pages} {30523} (\bibinfo {year} {2018})}\BibitemShut {NoStop}%
\bibitem [{\citenamefont {Xiao}\ \emph {et~al.}(2023)\citenamefont {Xiao}, \citenamefont {Liu}, \citenamefont {Wu}, \citenamefont {Peng},\ and\ \citenamefont {Guo}}]{xiao2023femtotesla}%
  \BibitemOpen
  \bibfield  {author} {\bibinfo {author} {\bibfnamefont {W.}~\bibnamefont {Xiao}}, \bibinfo {author} {\bibfnamefont {M.}~\bibnamefont {Liu}}, \bibinfo {author} {\bibfnamefont {T.}~\bibnamefont {Wu}}, \bibinfo {author} {\bibfnamefont {X.}~\bibnamefont {Peng}},\ and\ \bibinfo {author} {\bibfnamefont {H.}~\bibnamefont {Guo}},\ }\bibfield  {title} {\bibinfo {title} {Femtotesla atomic magnetometer employing diffusion optical pumping to search for exotic spin-dependent interactions},\ }\href {https://doi.org/10.1103/PhysRevLett.130.143201} {\bibfield  {journal} {\bibinfo  {journal} {Phys. Rev. Lett.}\ }\textbf {\bibinfo {volume} {130}},\ \bibinfo {pages} {143201} (\bibinfo {year} {2023})}\BibitemShut {NoStop}%
\bibitem [{\citenamefont {Cohen-Tannoudji}\ and\ \citenamefont {Dupont-Roc}(1972)}]{cohen1972experimental}%
  \BibitemOpen
  \bibfield  {author} {\bibinfo {author} {\bibfnamefont {C.}~\bibnamefont {Cohen-Tannoudji}}\ and\ \bibinfo {author} {\bibfnamefont {J.}~\bibnamefont {Dupont-Roc}},\ }\bibfield  {title} {\bibinfo {title} {Experimental study of zeeman light shifts in weak magnetic fields},\ }\href {https://doi.org/https://doi.org/10.1103/PhysRevA.5.968} {\bibfield  {journal} {\bibinfo  {journal} {Phys. Rev. A}\ }\textbf {\bibinfo {volume} {5}},\ \bibinfo {pages} {968} (\bibinfo {year} {1972})}\BibitemShut {NoStop}%
\bibitem [{\citenamefont {Zhang}\ \emph {et~al.}(2020{\natexlab{b}})\citenamefont {Zhang}, \citenamefont {Xiao}, \citenamefont {Ding}, \citenamefont {Feng}, \citenamefont {Peng}, \citenamefont {Shen}, \citenamefont {Sun}, \citenamefont {Wu}, \citenamefont {Wu}, \citenamefont {Yang} \emph {et~al.}}]{zhang2020recording}%
  \BibitemOpen
  \bibfield  {author} {\bibinfo {author} {\bibfnamefont {R.}~\bibnamefont {Zhang}}, \bibinfo {author} {\bibfnamefont {W.}~\bibnamefont {Xiao}}, \bibinfo {author} {\bibfnamefont {Y.}~\bibnamefont {Ding}}, \bibinfo {author} {\bibfnamefont {Y.}~\bibnamefont {Feng}}, \bibinfo {author} {\bibfnamefont {X.}~\bibnamefont {Peng}}, \bibinfo {author} {\bibfnamefont {L.}~\bibnamefont {Shen}}, \bibinfo {author} {\bibfnamefont {C.}~\bibnamefont {Sun}}, \bibinfo {author} {\bibfnamefont {T.}~\bibnamefont {Wu}}, \bibinfo {author} {\bibfnamefont {Y.}~\bibnamefont {Wu}}, \bibinfo {author} {\bibfnamefont {Y.}~\bibnamefont {Yang}}, \emph {et~al.},\ }\bibfield  {title} {\bibinfo {title} {Recording brain activities in unshielded \text{E}arth’s field with optically pumped atomic magnetometers},\ }\href {https://doi.org/10.1126/sciadv.aba8792} {\bibfield  {journal} {\bibinfo  {journal} {Sci. Adv.}\ }\textbf {\bibinfo {volume} {6}},\ \bibinfo {pages} {eaba8792} (\bibinfo {year} {2020}{\natexlab{b}})}\BibitemShut {NoStop}%
\bibitem [{\citenamefont {Wu}\ \emph {et~al.}(2024)\citenamefont {Wu}, \citenamefont {Zhang}, \citenamefont {Wei}, \citenamefont {Ou}, \citenamefont {Yue}, \citenamefont {Cheng},\ and\ \citenamefont {Liu}}]{wu2024effects}%
  \BibitemOpen
  \bibfield  {author} {\bibinfo {author} {\bibfnamefont {Z.}~\bibnamefont {Wu}}, \bibinfo {author} {\bibfnamefont {J.}~\bibnamefont {Zhang}}, \bibinfo {author} {\bibfnamefont {C.}~\bibnamefont {Wei}}, \bibinfo {author} {\bibfnamefont {Z.}~\bibnamefont {Ou}}, \bibinfo {author} {\bibfnamefont {H.}~\bibnamefont {Yue}}, \bibinfo {author} {\bibfnamefont {C.}~\bibnamefont {Cheng}},\ and\ \bibinfo {author} {\bibfnamefont {Y.}~\bibnamefont {Liu}},\ }\bibfield  {title} {\bibinfo {title} {Effects of system parameters on a single-beam synthetic gradiometer with a dual-cell structure},\ }\href {https://doi.org/https://doi.org/10.1364/OL.518697} {\bibfield  {journal} {\bibinfo  {journal} {Opt. Lett.}\ }\textbf {\bibinfo {volume} {49}},\ \bibinfo {pages} {2781} (\bibinfo {year} {2024})}\BibitemShut {NoStop}%
\bibitem [{\citenamefont {Li}\ \emph {et~al.}(2022)\citenamefont {Li}, \citenamefont {Liu}, \citenamefont {Jin}, \citenamefont {Akiti}, \citenamefont {Dai}, \citenamefont {Xu},\ and\ \citenamefont {Nwodom}}]{li2022kilohertz}%
  \BibitemOpen
  \bibfield  {author} {\bibinfo {author} {\bibfnamefont {S.}~\bibnamefont {Li}}, \bibinfo {author} {\bibfnamefont {J.}~\bibnamefont {Liu}}, \bibinfo {author} {\bibfnamefont {M.}~\bibnamefont {Jin}}, \bibinfo {author} {\bibfnamefont {K.~T.}\ \bibnamefont {Akiti}}, \bibinfo {author} {\bibfnamefont {P.}~\bibnamefont {Dai}}, \bibinfo {author} {\bibfnamefont {Z.}~\bibnamefont {Xu}},\ and\ \bibinfo {author} {\bibfnamefont {T.~E.-T.}\ \bibnamefont {Nwodom}},\ }\bibfield  {title} {\bibinfo {title} {A kilohertz bandwidth and sensitive scalar atomic magnetometer using an optical multipass cell},\ }\href {https://doi.org/https://doi.org/10.1016/j.measurement.2022.110704} {\bibfield  {journal} {\bibinfo  {journal} {Measurement}\ }\textbf {\bibinfo {volume} {190}},\ \bibinfo {pages} {110704} (\bibinfo {year} {2022})}\BibitemShut {NoStop}%
\bibitem [{\citenamefont {Yi}\ \emph {et~al.}(2024)\citenamefont {Yi}, \citenamefont {Liu}, \citenamefont {Wang}, \citenamefont {Xiao}, \citenamefont {Sheng}, \citenamefont {Peng},\ and\ \citenamefont {Guo}}]{yi2024free}%
  \BibitemOpen
  \bibfield  {author} {\bibinfo {author} {\bibfnamefont {K.}~\bibnamefont {Yi}}, \bibinfo {author} {\bibfnamefont {Y.}~\bibnamefont {Liu}}, \bibinfo {author} {\bibfnamefont {B.}~\bibnamefont {Wang}}, \bibinfo {author} {\bibfnamefont {W.}~\bibnamefont {Xiao}}, \bibinfo {author} {\bibfnamefont {D.}~\bibnamefont {Sheng}}, \bibinfo {author} {\bibfnamefont {X.}~\bibnamefont {Peng}},\ and\ \bibinfo {author} {\bibfnamefont {H.}~\bibnamefont {Guo}},\ }\bibfield  {title} {\bibinfo {title} {Free-induction-decay $^4\text{He}$ magnetometer using a multipass cell},\ }\href {https://doi.org/10.1103/PhysRevApplied.22.014084} {\bibfield  {journal} {\bibinfo  {journal} {Phys. Rev. Appl.}\ }\textbf {\bibinfo {volume} {22}},\ \bibinfo {pages} {014084} (\bibinfo {year} {2024})}\BibitemShut {NoStop}%
\bibitem [{\citenamefont {Afach}\ \emph {et~al.}(2024)\citenamefont {Afach}, \citenamefont {Aybas~Tumturk}, \citenamefont {Bekker}, \citenamefont {Buchler}, \citenamefont {Budker}, \citenamefont {Cervantes}, \citenamefont {Derevianko}, \citenamefont {Eby}, \citenamefont {Figueroa}, \citenamefont {Folman} \emph {et~al.}}]{afach2024can}%
  \BibitemOpen
  \bibfield  {author} {\bibinfo {author} {\bibfnamefont {S.}~\bibnamefont {Afach}}, \bibinfo {author} {\bibfnamefont {D.}~\bibnamefont {Aybas~Tumturk}}, \bibinfo {author} {\bibfnamefont {H.}~\bibnamefont {Bekker}}, \bibinfo {author} {\bibfnamefont {B.~C.}\ \bibnamefont {Buchler}}, \bibinfo {author} {\bibfnamefont {D.}~\bibnamefont {Budker}}, \bibinfo {author} {\bibfnamefont {K.}~\bibnamefont {Cervantes}}, \bibinfo {author} {\bibfnamefont {A.}~\bibnamefont {Derevianko}}, \bibinfo {author} {\bibfnamefont {J.}~\bibnamefont {Eby}}, \bibinfo {author} {\bibfnamefont {N.~L.}\ \bibnamefont {Figueroa}}, \bibinfo {author} {\bibfnamefont {R.}~\bibnamefont {Folman}}, \emph {et~al.},\ }\bibfield  {title} {\bibinfo {title} {What can a \text{GNOME} do? \text{Search} targets for the global network of optical magnetometers for exotic physics searches},\ }\href {https://doi.org/https://doi.org/10.1002/andp.202300083} {\bibfield  {journal} {\bibinfo  {journal} {Ann. Phys. (Berlin)}\ }\textbf {\bibinfo {volume} {536}},\ \bibinfo
  {pages} {2300083} (\bibinfo {year} {2024})}\BibitemShut {NoStop}%
\end{thebibliography}

%

\end{document}